\documentclass[journal ]{new-aiaa}
\usepackage[utf8]{inputenc}
\usepackage{textcomp}
\usepackage{graphicx}
\usepackage{amsmath}
\usepackage{float}
\usepackage{subfigure}
\usepackage{multirow}
\usepackage{subcaption}
\usepackage{tabularx}
\usepackage{booktabs}
\usepackage{arydshln}
\usepackage{bm}
\usepackage[version=4]{mhchem}
\usepackage{siunitx}
\usepackage{longtable,tabularx}
\newcommand{\vect}[1]{\mathbf{#1}}
\newcommand{\matr}[1]{\mathbf{#1}} 
\usepackage{eso-pic}
\AddToShipoutPictureBG{%
 \AtPageLowerLeft{%
  \raisebox{3\baselineskip}{\makebox[\paperwidth]{\begin{minipage}{21cm}\centering
  \emph{\textbf{NAVAIR Public Release 2025-0629. Distribution Statement A - "Approved for public release; distribution is unlimited"}}
  \end{minipage}}}
  }
  }
\title{Bandwidth of Linear Classically Damped Systems with Application to Experimental Model Aircraft}

\author{Benjamin J. Chang \footnote{Graduate Student, Department of Aerospace Engineering; also, Flutter and Dynamics Engineer, Naval Air Warfare Center Aircraft Division}}
\affil{University of Illinois Urbana-Champaign, Urbana, Illinois, 61801}

\author{Keegan J. Moore\footnote{Associate Professor, Daniel Guggenheim School of Aerospace Engineering}}
\affil{Georgia Institute of Technology, Atlanta, Georgia, 30332}

\author{Lawrence A. Bergman\footnote{Professor Emeritus and Research Professor, Department of Aerospace Engineering, Associate AIAA Fellow.}}
\affil{University of Illinois Urbana-Champaign, Urbana, Illinois, 61801}

\author{Alexander F. Vakakis\footnote{Professor, Department of Mechanical Science and Engineering.}}
\affil{University of Illinois Urbana-Champaign, Urbana, Illinois, 61801}

\author{Walter A. Silva\footnote{Senior Research Scientist, Aeroelasticity Branch, AIAA Fellow.}}
\affil{NASA Langley Research Center, Hampton, Virginia, 23681}

\begin{document}

\maketitle

%-----------------------------------------------------------------
%% ABSTRACT-------------------------------------------------
\begin{abstract}
Bandwidth is a widely known concept and tool used in structural dynamics to measure an oscillator's capacity to dissipate energy over time, for example when used in half-power damping estimation of structural modes. Root Mean Square (RMS) Bandwidth is a generalization of bandwidth that overcomes some of the limitations encountered with conventional bandwidth, including the prerequisite of linearity, single-mode response, and light damping. However, its mathematical form does not reveal much about the physics behind it. In this paper, we extend RMS Bandwidth to multiple degree-of-freedom, linear, time-invariant, classically damped systems by deriving an Analytical Root Mean Square (ARMS) Bandwidth in terms of a system's modal parameters and initial modal energy distribution. We demonstrate that ARMS Bandwidth reliably and accurately computes a single measure for a practical structure's dissipative capacity. Also, a purely data-driven methodology for assessing the modal energy distribution is developed. We apply ARMS Bandwidth to single and multiple degree-of-freedom systems and an experimental model aircraft to demonstrate its broad applicability. Future work will address the effects of non-classical damping distribution, time-varying parameters, and nonlinearities.
\end{abstract}

%-----------------------------------------------------------------
%% INTRODUCTION -------------------------------------------
\section{Introduction}
\lettrine{B}{andwidth} is a concept accepted and used across a wide array of science and engineering disciplines. However, its definition and mathematical form depends on the field and specific application. In control systems, bandwidth is typically quantified in the frequency domain and is a performance metric for filters, operational amplifiers, and actuators. Bandpass filters, such as those used in adaptive control of aircraft engine stall modes, permit signals within the passband, generally quantified by half-power bandwidth, to continue while attenuating all other signals \cite{NiseControls,CompStall}. The bandwidth of liquid fuel flow actuators quantifies the frequency range over which the device is effective \cite{fuelflow,fuelflow2}. In antenna applications, the bandwidth of zero reactance tuned antennas is quantified in terms of conductance bandwidth, the frequency range over which the antenna accepts power, and matched voltage-standing-wave-ratio bandwidth, the frequency range associated with the constant magnitude of the reflection coefficient squared \cite{antennaQ}. From a practical standpoint, the latter quantifies impedance matching for the antenna which affects performance. In these cases, among many others, bandwidth involves a frequency band over which the performance of a system can be quantified.

\begin{figure}[H]
\centering
\includegraphics[width=0.6\textwidth]{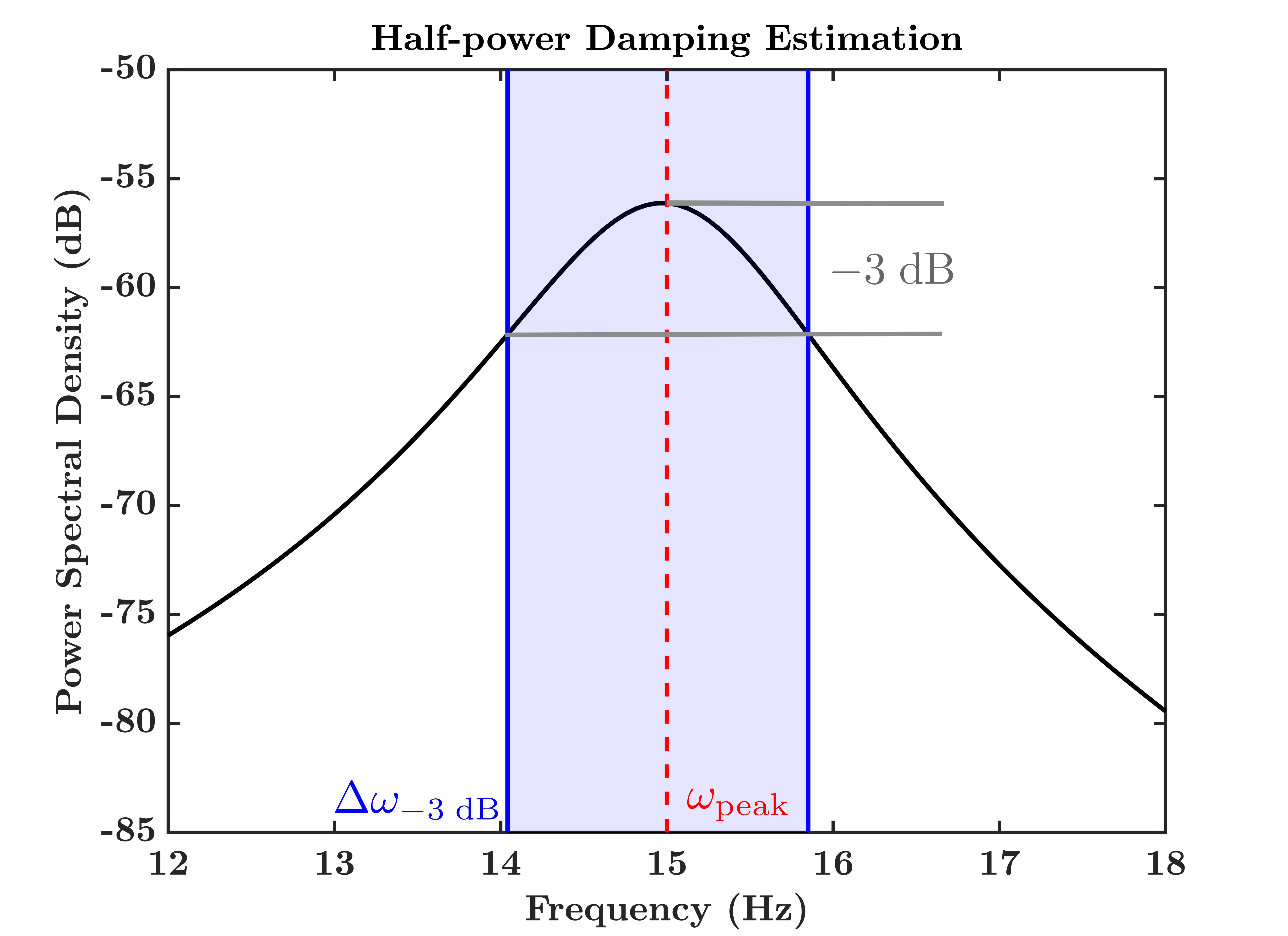}
\caption{Demonstration of half-power bandwidth used in damping estimation of a monochromatic response.}
\label{fig:HPBW}
\end{figure}

Considering a single degree-of-freedom (SDOF) oscillator in the context of traditional linear structural dynamics and modal analysis, the half-power bandwidth $\Delta\omega_{\text{-3\;dB}}$ is used to compute the quality factor $Q$ as a means to estimate the viscous modal damping ratio through the relation $Q=\frac{\omega_{\text{peak}}}{\Delta\omega_{-3\;\text{dB}}}=\frac{1}{2\zeta}$. Here, $\Delta\omega_{-3\;\text{dB}}$ is the frequency band over which the amplitude is 3 dB below the amplitude at resonance, i.e., the frequency band over which the power at resonance is halved \cite{HPBW,HPBW2}. For example, for a response in the form of a decaying sine wave of frequency $\omega$ and damping ratio $\zeta$, i.e., $e^{-\zeta\omega t}\sin(\omega t)$, the power spectral density function can be used to obtain the half-power bandwidth and from it the damping ratio as visualized in Fig. \ref{fig:HPBW}. The classical definition of bandwidth is used to quantify the distribution of the energy of a structural mode in the frequency domain, or equivalently, the dissipative capacity of that mode. Currently, there is no similar scalar measure that quantifies the energy distribution in frequency of a multiple degree-of-freedom (MDOF) oscillating mechanical system or any other practical engineering structure, or equivalently its overall dissipative capacity. In other words, whereas one can compute a "local" bandwidth measure for a single well-separated structural mode, there is no similar method for computing the "global" bandwidth of a multi-modal response, and thus assess the overall dissipative capacity of an entire MDOF structure. This foundational limitation is addressed in the present work.
Mojahed et. al. generalized the concept of (half-power) bandwidth via modification of the Root Mean Square (RMS) Bandwidth expression that is commonly used in electrical engineering and signal processing applications \cite{jsv2022alireza,sigproc2014}. For an SDOF oscillator, the modified form of RMS Bandwidth is defined as,
\begin{equation}
\label{sample:RMSBW}
\Delta\omega_{\text{RMS}} = 2\sqrt{\frac{\displaystyle\int_{0}^{\infty}\omega^{2}|V_{E}(\omega)|^{4}d\omega}{\displaystyle\int_{0}^{\infty}|V_{E}(\omega)|^{4}d\omega}},
\end{equation}
where $V_{E}(\omega)$ is the Fourier transform of the \emph{envelope} of the velocity time series, $V_{E}(\omega) = \displaystyle\int_{-\infty}^{\infty} v_{E}(t) e^{-j\omega t} dt$. Note that Eq.~\eqref{sample:RMSBW} has units of cyclic frequency (rad/s) but this (as well as other physical units) will be omitted in the following discussion for convenience. This expression assumes that the system's velocity envelope monotonically decays and contains minimal frequency content.  If the velocity envelope does not monotonically decay, then $\omega^{2}$ is replaced with $(\omega - \omega_{c})^{2}$, where $\omega_{c}$ is the frequency corresponding to the maximum of the Fourier transform of the envelope and is called the central frequency. However, in a passive SDOF mechanical oscillator during a free decay, it always holds that $\omega_{c}=0$, i.e., the velocity envelope monotonically decays. Mojahed et. al. demonstrated numerically that $\Delta\omega_{\text{RMS}}$ recovered the modal damping coefficient for a Duffing oscillator at low energies, i.e., when the response was essentially linear. This tied RMS Bandwidth to the dissipative capacity of an oscillator, i.e., to its ability to dissipate energy over time.

This work builds on RMS Bandwidth by deriving an Analytical Root Mean Square (ARMS) Bandwidth expression that can be used to quantify the overall dissipative capacity of a multi-modal structure by directly relating $\Delta\omega_{\text{RMS}}$ to the initial modal energy distribution and modal parameters of a structure under an impulsive load. We provide application of ARMS Bandwidth to SDOF systems, MDOF systems, and a reduced-order model (ROM) of an experimental model aircraft. In the process, we discuss issues related to the numerical computation of ARMS Bandwidth to ensure accuracy and ease of use in its practical implementation to real-life engineering problems.

%-----------------------------------------------------------------
%% METHODS OF ANALYSIS -------------------------------------------
\section{Methods of Analysis}
We first discuss fundamental concepts in frequency and time-frequency analysis relevant to the computation of ARMS Bandwidth.

\subsection{Theoretical Foundations}
\subsubsection{Fourier Transform}
The Fourier transform is a linear integral transform that takes a stationary time signal $f(t)$ and outputs its frequency-domain representation $F(\omega)$ defined as
\begin{equation}
\label{sample:FT}
F(\omega) = \displaystyle\int_{-\infty}^{\infty} f(t) e^{-j\omega t} dt,
\end{equation}
\begin{equation}
\label{sample:DFT}
X(k) = \displaystyle\sum_{n=1}^{N} x(n)e^{-(\frac{-j2\pi}{N})(n-1)(k-1)}.
\end{equation}
The most common numerical implementation of the discrete Fourier transform of a time series $x$ shown in Eq.~\eqref{sample:DFT} is the fast Fourier transform (FFT), which is typically coded as an N-point FFT or arbitrary-point FFT. The former zero-pads or truncates the signal such that its length is a power of two, i.e., $N_{\text{samples}} = 2^{n}$, while the latter does not require any manipulation of signal length. The computation of both RMS and ARMS Bandwidth, as shown later, is affected by user selection of FFT parameters.

\subsubsection{Wavelet Transform}
The wavelet transform is an integral transform that provides a time-frequency representation of a stationary or nonstationary signal. The continuous wavelet transform is the convolution of a mother wavelet function and the signal of interest $z(\tau)$ defined as
\begin{equation}
\label{sample:contWT}
Z(\omega,t) = \displaystyle\int_{-\infty}^{\infty} z(\tau)\psi_{\omega,t}^{*}(\tau) d\tau,
\end{equation}
where $(\cdot)^{*}$ denotes the complex conjugate and the function $\psi_{\omega,t}(\tau)$ is the normalized Morlet mother wavelet defined as
\begin{equation}
\label{sample:morlet}
\psi_{\omega,t}(\tau) = \pi^{-\frac{1}{4}} e^{j\omega_{0}\Big( \frac{\tau-t}{\omega}\Big)}e^{-\frac{1}{2}\Big( \frac{\tau-t}{\omega}\Big)^{2}},
\end{equation}
with $\omega_{0}$ being the center frequency. The maximum wavelet transform is defined as
\begin{equation}
\label{sample:MaxWavelet}
\tilde{Z}(\omega,t) = \max_{t} \Bigg[\displaystyle\int_{-\infty}^{\infty} z(\tau)\psi_{\omega,t}^{*}(\tau) d\tau \Bigg],
\end{equation}
which is two-dimensional like a Fourier transform, but can account for time-frequency effects, such as nonlinear hardening or vibro-impacts \cite{mooreWBEMD,mooreWBEMD2}. The maximum wavelet transform is used in the data-driven methodology that will be outlined in a later section for computing ARMS Bandwidth. Indeed the main advantage of the wavelet transform compared to the Fourier transform is that it is applicable to stationary as well as nonstationary signals (e.g., nonlinear responses, responses of time-varying systems, or responses with closely spaced modes), and that it computes the temporal evolution of the dominant harmonic components in a measured time series. As mentioned, this last feature will prove to be important for the data-driven methodology that will be developed in a later section for ARMS Bandwidth computation.

\subsection{ARMS Bandwidth}
The RMS Bandwidth of a linear, time-invariant SDOF system can be calculated directly by enveloping its velocity time history and evaluating Eq.~\eqref{sample:RMSBW} numerically. Mojahed et. al. demonstrated that for the Duffing oscillator at low energy levels (so that the response is nearly linear), the RMS Bandwidth computed via Eq.~\eqref{sample:RMSBW} is equal to the modal damping coefficient \cite{jsv2022alireza}. However, this was only demonstrated numerically. In addition, the requirement of enveloping introduces the numerical challenge of automating the envelope generation process, e.g., specifying local maxima spacing between Akima spline points \cite{akima}. To overcome challenges with computing RMS Bandwidth and to uncover the underlying dynamics behind it, we establish an analytical, closed-form relationship between RMS Bandwidth, initial modal energy distribution, and modal parameters of generic linear systems. We designate the resulting expression as ARMS Bandwidth. In the next sections we derive expressions for the ARMS Bandwidth of linear SDOF oscillators and classically damped linear MDOF oscillators.

\subsubsection{Single Degree-of-Freedom ARMS Bandwidth}
Consider a linear SDOF oscillator with natural frequency $\omega$ and viscous damping ratio $\zeta$ determined by its mass $m$, stiffness coefficient $k$, and damping coefficient $c$. Given an impulsive excitation, equivalent to non-zero velocity and zero displacement initial conditions, the resulting response will be an exponentially decaying sinusoid. Note that the modal response of a single (well-separated) mode of a linear, classically damped MDOF system can also be treated as the response of an equivalent SDOF oscillator, since no modal interactions (energy exchanges) can occur and the equations of motion can be decoupled. For an SDOF oscillator, assume the corresponding envelope of the velocity time series during free decay is in the following form,
\begin{equation}
\label{sample:SDOFX}
v_{E}^{(n)}(t) = C e^{-\zeta \omega t},
\end{equation}
where $C$ is the amplitude of the modal velocity. To calculate the RMS Bandwidth of the oscillator using Eq.~\eqref{sample:RMSBW}, we determine the Fourier transform of Eq.~\eqref{sample:SDOFX}, compute its modulus to the fourth power, and lastly analytically evaluate each integral. As shown in the Appendix, in this case the RMS Bandwidth, Eq.~\eqref{sample:RMSBW}, yields the well-known half-power bandwidth as defined in classical vibration theory:
\begin{equation}
\label{sample:SDOFfinal}
\Delta\omega_{\text{ARMS}} = 2\sqrt{\frac{\displaystyle\int_{0}^{\infty}\omega^{2}|V_{E}(\omega)|^{4}d\omega}{\displaystyle\int_{0}^{\infty}|V_{E}(\omega)|^{4}d\omega}} = 2\sqrt{\frac{\frac{\pi}{4}\frac{1}{\zeta\omega}}{\frac{\pi}{4}\frac{1}{\zeta^{3}\omega^{3}}}}= 2\zeta\omega.
\end{equation}
As shown in Eq.~\eqref{sample:SDOFfinal}, the ARMS Bandwidth that is derived from a velocity time series corresponding to a single mode is the \emph{modal damping coefficient} solely comprised of natural frequency and damping ratio, i.e., the classical bandwidth definition of the single mode. By inspection, if the kinetic energy envelope is used in lieu of the velocity envelope, we recover twice the modal damping coefficient as the ARMS Bandwidth. In addition, the amplitude of the velocity envelope cancels out in the computation, demonstrating that (i) ARMS Bandwidth for an SDOF system is energy invariant (as expected due to the linearity of the mode) and (ii) displacement, velocity, or acceleration free decays may be used in the ARMS Bandwidth computation.

\subsubsection{Multiple Degree-of-Freedom ARMS Bandwidth}
While one can envelope the velocity time series of any degree-of-freedom (DOF) of an MDOF system to compute RMS Bandwidth, the physical meaning of the obtained quantity is not immediately clear. However, ARMS Bandwidth is physics-based and has a closed-form expression valid for systems that satisfy the requirements to compute ARMS Bandwidth, so we can use ARMS Bandwidth to find physical meaning. The ultimate question is \emph{"if the ARMS Bandwidth of a single mode is the modal damping coefficient, how does each mode in the multi-modal response of an MDOF system contribute to the overall ARMS Bandwidth of the system?"}. To this end, considering a general mechanical $N$-DOF linear dynamical system, assume that the envelope of the velocity $v_{E}$ of a specific DOF at a specific location $\vect{r}=x\hat{i}_{1} + y\hat{i}_{2} + z\hat{i}_{3}$ along the specific axis (direction) $\hat{i}_{j}$ is comprised of $N$ modal (SDOF-equivalent) responses of the same form as Eq.~\eqref{sample:SDOFX},
\begin{equation}
\label{sample:MDOFX}
v_{E}(\vect{r},\hat{i}_{j},t) = \displaystyle\sum_{n=1}^{N} v_{E}^{(n)}(t) = \displaystyle\sum_{n=1}^{N} C_{n} e^{-b_{n} t}, \;\;\; b_{n}=\zeta_{n} \omega_{n},
\end{equation}
where $C_{n}$ represents the corresponding initial amplitude of the modal velocity envelope (modal velocity participation) of the $n^{\text{th}}$ mode within the frequency range of interest. The assumption of exponential decay of each modal response is valid for a classical viscous damping distribution, where no modal coupling exists and all modes are real-valued. It doesn't hold, however, for a non-classical viscous damping distribution when the modes become complex and modal interactions occur \cite{Ewins}. By inspection, the Fourier transform of each term in the series is identical in form to $V_{E}(\omega)$ in the SDOF case, and given that the Fourier transform is a linear integral operation, it holds that:
\begin{equation}
\label{sample:MDOFFFT}
V_{E}(\omega) = \displaystyle\sum_{n=1}^{N} C_{n} \frac{\zeta_{n}\omega_{n} - j\omega}{\zeta_{n}^{2}\omega_{n}^{2} + \omega^{2}}.
\end{equation}
Evaluation of $|V_{E}(\omega)|$ is intractable from a practical point of view in this form, so we need to convert this complex summation to polar form such that $r_{n}=\frac{C_{n}}{\sqrt{b_{n}^{2} + \omega^{2}}}$ and $\theta_{n}=\tan^{-1}\Big( \frac{-\omega}{b_{n}}\Big)$ to obtain:
\begin{equation}
\label{sample:MDOFpolar}
\resizebox{.9 \textwidth}{!} 
{
$
|V_{E}(\omega)|^{2}= \Biggl|\displaystyle\sum_{n=1}^{N} r_{n} e^{j\theta_{n}}\Biggr|^{2} = \Biggl|\displaystyle\sum_{n=1}^{N} \Big( r_{n}\cos(\theta_{n})\Big) + j\displaystyle\sum_{n=1}^{N} \Big(  r_{n}\sin(\theta_{n}\Big)\Biggr|^{2} = \displaystyle\sum_{n=1}^{N}\displaystyle\sum_{m=1}^{N} r_{n}r_{m}\Big(\cos(\theta_{n})\cos(\theta_{m}) + \sin(\theta_{n})\sin(\theta_{m})\Big)
$
}.
\end{equation}
Recognizing that $\cos(\theta_{n}) = \frac{b_{n}}{\sqrt{b_{n}^{2} + \omega_{n}^{2}}}$ and $\sin(\theta_{n}) = \frac{\omega_{n}}{\sqrt{b_{n}^{2} + \omega_{n}^{2}}}$, we employ Euler's formula to obtain:
\begin{equation}
\label{sample:MDOFpolar2}
|V_{E}(\omega)|^{4}= \displaystyle\sum_{n=1}^{N}\displaystyle\sum_{m=1}^{N}\displaystyle\sum_{i=1}^{N}\displaystyle\sum_{j=1}^{N} \Big( 
\frac{C_{n}C_{m}C_{i}C_{j}(b_{n}b_{m}+\omega^{2})(b_{i}b_{j}+\omega^{2})}{(b_{n}^2 + \omega^{2})(b_{m}^2 + \omega^{2})(b_{i}^2 + \omega^{2})(b_{j}^2 + \omega^{2})}\Big).
\end{equation}
Each integral in the RMS Bandwidth formula, Eq.~\eqref{sample:RMSBW}, can now be evaluated via symbolic algebra to obtain an exact analytical expression for the ARMS Bandwidth of a linear, classically damped MDOF system:
\begin{equation}
\label{sample:ARMSBWFINAL}
\resizebox{.9 \textwidth}{!} 
{
$
\Delta\omega_{\text{ARMS}} =  
2\sqrt{\frac{ \displaystyle\sum_{n=1}^{N}\displaystyle\sum_{m=1}^{N}\displaystyle\sum_{i=1}^{N}\displaystyle\sum_{j=1}^{N} C_{n}C_{m}C_{i}C_{j} \frac{(b_{n}b_{j}(b_{m}+b_{i}) + b_{n}b_{m}b_{i} + b_{m}b_{i}b_{j})(b_{n}(2b_{m}+b_{i}+b_{j}) + b_{m}(b_{i} + b_{j}) + 2b_{i}b_{j})}{(b_{n}+b_{m})(b_{n}+b_{i})(b_{n}+b_{j})(b_{m}+b_{i})(b_{m}+b_{j})(b_{i}+b_{j})}}{\displaystyle\sum_{n=1}^{N}\displaystyle\sum_{m=1}^{N}\displaystyle\sum_{i=1}^{N}\displaystyle\sum_{j=1}^{N} C_{n}C_{m}C_{i}C_{j} \frac{(b_{n}+b_{m}+b_{i}+b_{j})(b_{n}(2b_{m}+b_{i}+b_{j}) + b_{m}(b_{i} + b_{j}) + 2b_{i}b_{j})}{(b_{n}+b_{m})(b_{n}+b_{i})(b_{n}+b_{j})(b_{m}+b_{i})(b_{m}+b_{j})(b_{i}+b_{j})}}}
$}.
\end{equation}
\emph{This expression computes the bandwidth of any linear, classically damped multi-modal system in terms of its modal properties and initial modal energy distribution}. Specifically, it shows that the ARMS Bandwidth computed from the free decay velocity time series $v_{E}(\vect{r},\hat{i}_{j},t)$ is exclusively a function of the modal parameters of the system captured by terms $b_{n} = \omega_{n}\zeta_{n}$ and the initial modal energy distribution captured by the modal velocity participations $C_{n}$. In terms of the dynamics, \emph{the ARMS Bandwidth provides a quantification of the overall dissipative capacity of the structure when excited by an impulse at the location and direction considered}, as it is comprised of modal dissipative capacities (modal damping coefficients) weighted by the initial modal energy distribution, i.e., the initial modal energies following the application of the impulse. ARMS Bandwidth is also tied to how the impulsive force input or initial conditions, e.g., different pulse durations, affect the modal energy distribution summarized by $C_{n}$. A study of this is presented in a subsequent section. It should be noted that Eq.~\eqref{sample:ARMSBWFINAL} recovers ARMS Bandwidth for SDOF systems when $N=1$. 

When applied to a single modal velocity, the ARMS Bandwidth is a measure of that mode's dissipative capacity. When applied to the velocity response at a specific location and along a certain direction of a general MDOF system, ARMS Bandwidth is a measure of the \emph{local dissipative capacity} of the system. One may compute the local ARMS Bandwidth based on the structural velocity at the point of application and direction of the impulsive excitation, similar to the analogous computation for an SDOF system where the point of application and the direction of the excitation are uniquely determined, or at non-drive point locations but along the same direction of impulse application. While Eq.~\eqref{sample:ARMSBWFINAL} is tedious, it can be compactly implemented in code as four nested loops. Subsequent sections describe a purely data-driven method to (i) compute ARMS Bandwidth directly from acquired modal test data, and (ii) visualize the spatial dissipative capacity of a system by applying impulsive forces at varying locations on the structure. Ultimately, the ARMS Bandwidth characterizes the dissipative capacity of a linear oscillator (mode), or a multi-modal structure when impulsively forced at a certain location and along a certain direction. In the next sections, we provide applications of ARMS Bandwidth to numerical SDOF and MDOF systems as well as an experimental model aircraft.

In synopsis, for an MDOF linear, time-invariant, classically damped system, e.g., the ROM of an aircraft, we may define two types of bandwidth measures that quantify its dissipative capacity. The \emph{local dissipative} capacity of the ROM is assessed by computing the ARMS Bandwidth using the velocity at a single location along a single direction, which depends on the modal participations at that point along that axis. For example, energy may be dissipated faster in an aircraft when an attachment (store) is placed on the inboard wing compared to the outboard wing (see results in a later section). The \emph{spatially distributed} dissipative capacity of the ROM is the local ARMS Bandwidth computed at various locations that form a three-dimensional visualization of the dissipative capacity of the entire system given a specific impulse at a specific location. For example, this measure could be used to compare different structural layouts of an aircraft to maximize the dissipative capacity of the aircraft within specific regions, say to reduce the overall dynamic response of the wing under a specific loading condition like ejecting a wing store, which would impart an impulsive load to the wing.

%-----------------------------------------------------------------
%% APPLICATION OF BANDWIDTH TO SDOF SYSTEMS-----------------------------------
\section{Application of ARMS Bandwidth to SDOF Systems}
In this section, we provide a first validation of the ARMS Bandwidth expression in Eq.~\eqref{sample:ARMSBWFINAL} for a linear, time-invariant SDOF system, a linear, time-varying SDOF system, and a nonlinear Duffing oscillator at low energy levels. In addition, we make important observations regarding the influence of data processing choices on the RMS Bandwidth computation.

\subsection{Linear Time-Invariant SDOF System}
\begin{figure}[H]
\centering
\includegraphics[width=0.16\textwidth]{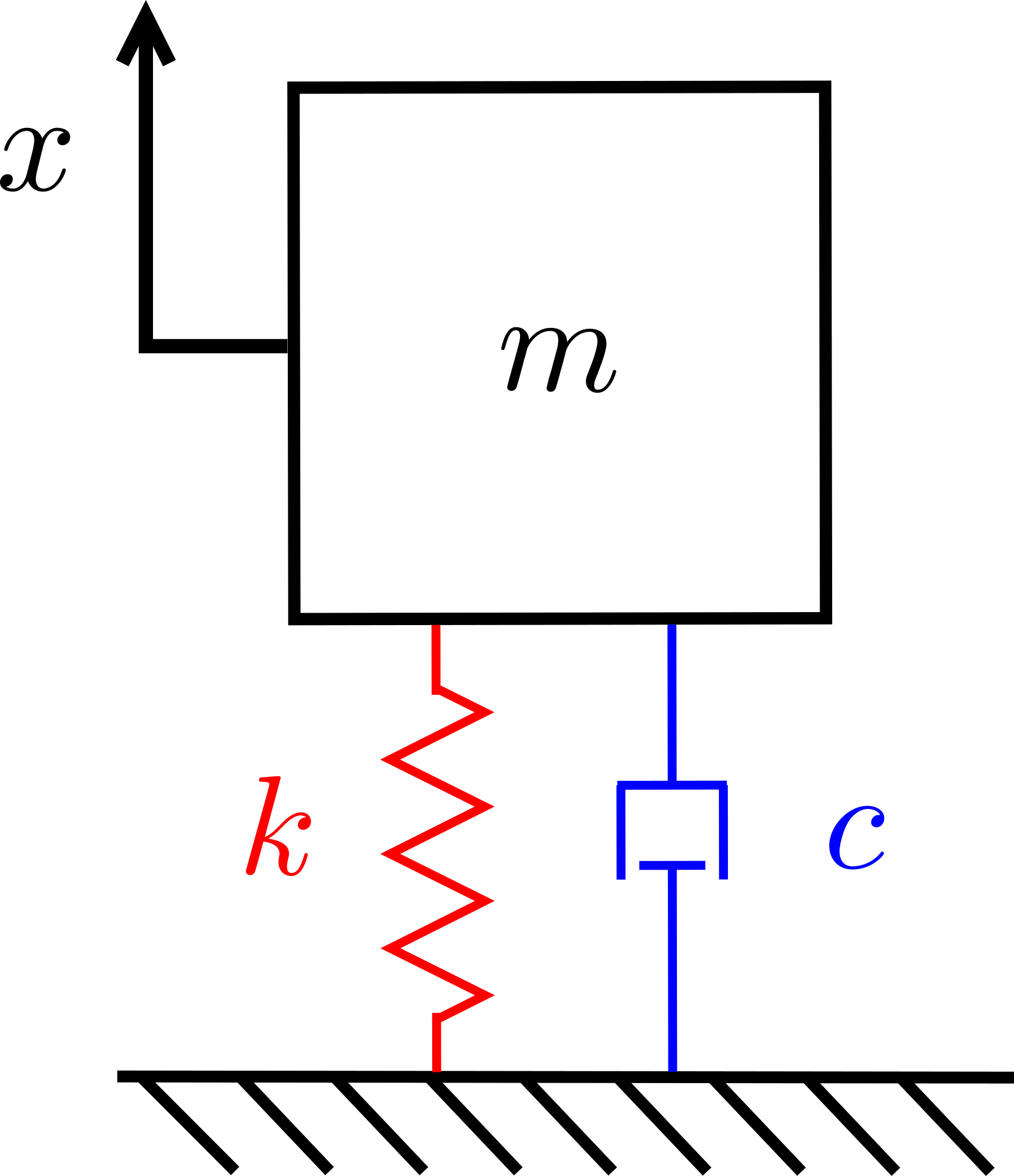}
\caption{Linear, time-invariant SDOF system.}
\label{fig:SDOFsys}
\end{figure}

We start by investigating the ARMS Bandwidth of a linear, time-invariant SDOF system of the form $m\ddot{x}+c\dot{x}+kx=0$ with $m=4$, $k=50$, and variable physical damping coefficient $c\in [0.4,0.8]$ shown in Fig. \ref{fig:SDOFsys}. Note that for the sake of convenience we omit units for the different system parameters from here on unless otherwise noted. The system is given initial conditions of $[x(0),\dot{x}(0)]=[0,1]$ and is integrated for 80 and 150 seconds with $\Delta t=0.0032$ second using a 4\textsuperscript{th}-order Runge-Kutta integration scheme for each value of physical damping coefficient. These two simulation durations were chosen to demonstrate the effect of signal decay on the bandwidth computation. For the numerical RMS Bandwidth, Akima spline interpolation is used to generate the velocity envelope. Alternatively, for ARMS Bandwidth, the maximum value of the velocity envelope $C$ (which is trivial in the case of an SDOF oscillator), natural frequency $\omega=\sqrt{k/m}$, and damping ratio $\zeta = c/2m\omega$ were used with Eq.~\eqref{sample:ARMSBWFINAL}. 

\begin{figure}[H]
\centering
\includegraphics[width=1\textwidth]{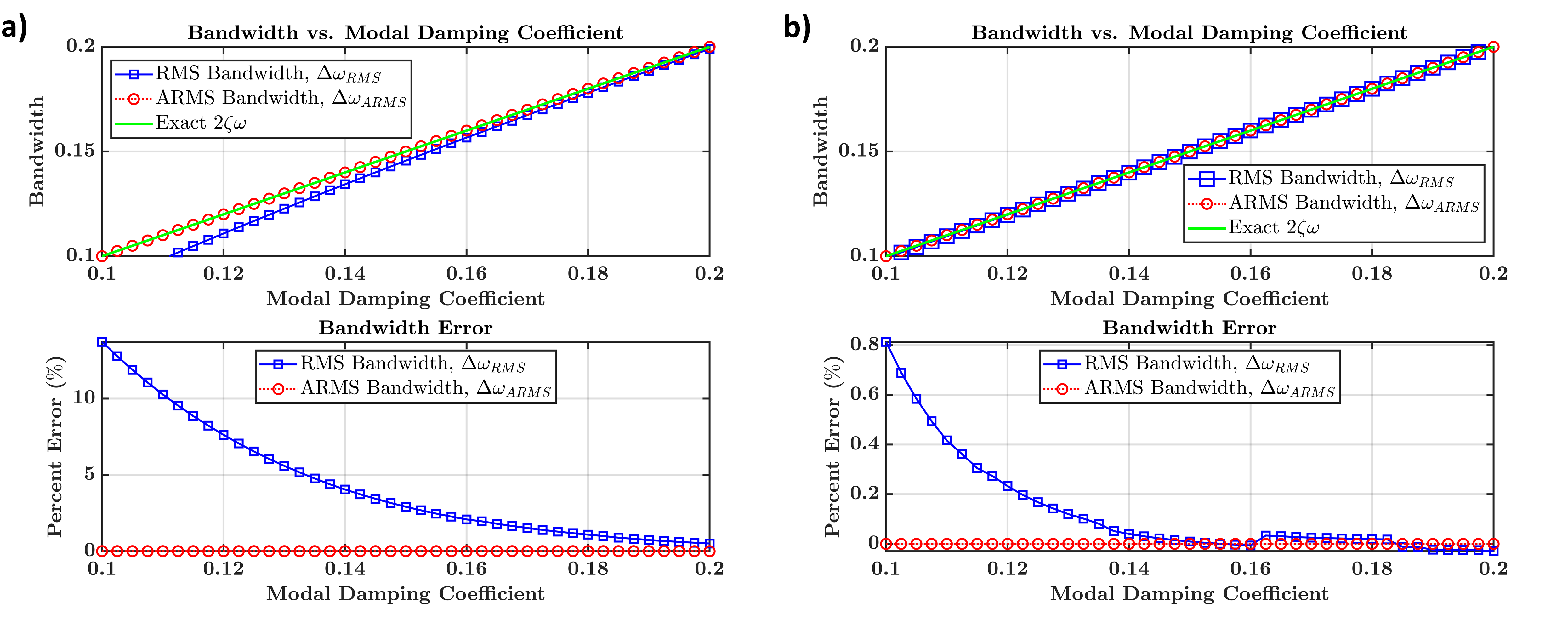}
\caption{Comparison of RMS and ARMS Bandwidth results using different simulation times: a) 80-second simulation results and b) 150-second simulation results.}
\label{fig:SDOFstudy}
\end{figure}

Comparisons of computed RMS and ARMS Bandwidth to the exact bandwidth of a lightly damped, i.e., $\zeta << 1$, SDOF system, $\Delta\omega_{\text{exact}}=2\zeta\omega$, for both groups of simulations are presented in Fig. \ref{fig:SDOFstudy}. In addition, the maximum percent error in the computed bandwidth for each group of simulations is provided in Table \ref{tab:error}. One notes that the RMS Bandwidth error is non-trivial in both cases because its computation uses the numerical FFT of the entire velocity envelope, which (i) is sensitive to envelope generation parameters and (ii) requires a sufficiently long decay time window for convergence. The final-to-initial velocity amplitude ratios for the 80-second simulation are $1.83$\% and $3.36\text{e-}2$\% for the lowest and greatest damping ratios, respectively. The final-to-initial velocity amplitude ratios for the 150-second simulation are $4.41\text{e-}2$\% and $2\text{e-}6$\% for the lowest and greatest damping ratios, respectively. Increasing the simulation time such that the signal amplitude damps out significantly greatly increases accuracy for RMS Bandwidth, but requires additional computation time. However, the ARMS Bandwidth error is trivial across all simulation durations and system damping ratios, as its computation only requires the maximum envelope amplitude and the modal parameters, both of which are known and unchanging. While both bandwidth computations yield the modal damping coefficient as expected, these results demonstrate the reduced simulation time (or experimental time history length in real measurements) required for accurate computation of ARMS Bandwidth.
\begin{table}[htbp]
  \caption{Simulation time and maximum bandwidth error for each simulation group.}
\makebox[\textwidth][c]{
    \begin{tabular}{ccc}
\cmidrule(lr){1-3}
Simulation Time (s) & $err(\Delta\omega_{\text{RMS}})$  (\%) & $err(\Delta\omega_{\text{ARMS}})$  (\%)  \\ \midrule
80 & $13.7$ & $10^{-14}$ \\
150 & $0.8$ & $10^{-14}$ \\
\bottomrule
\end{tabular}}
\label{tab:error}%
\end{table}%

\subsection{Linear Time-Varying SDOF System}
\begin{figure}[H]
\centering
\includegraphics[width=1\textwidth]{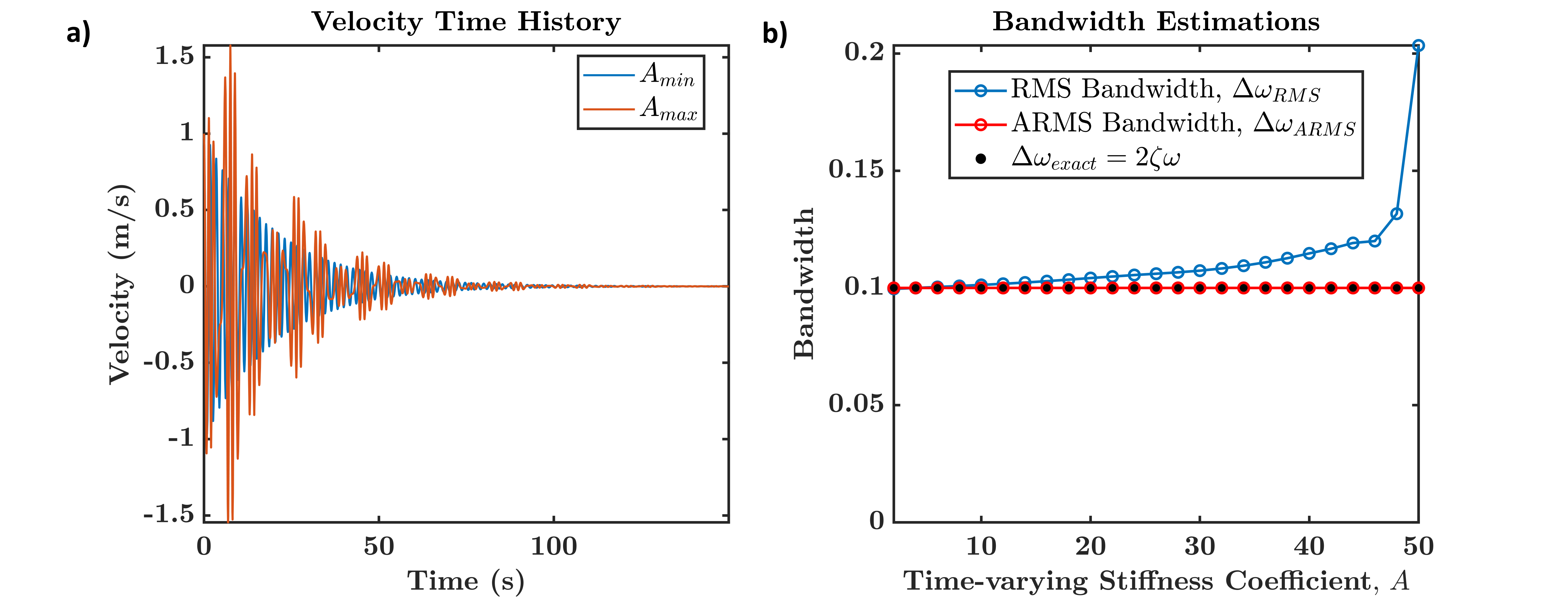}
\caption{Comparison of simulation results using extreme values ($A_{min}=5$ and $A_{max}=50$) of the time-varying stiffness coefficient: a) velocity time series and b) RMS and ARMS Bandwidth.}
\label{fig:TV}
\end{figure}

To demonstrate the broad applicability of the expression for ARMS Bandwidth, we investigate the bandwidth of a time-varying SDOF system of the form $m\ddot{x}+c\dot{x}+(k+A\sin(t))x=0$ with $m=4$, $k=50$, $c=0.40$, and variable time-varying stiffness coefficient $A\in [5,50]$. While both natural frequency and damping ratio vary with time as shown in Eq.~\eqref{sample:LTVsystem}, the modal damping coefficient is constant, $c_{m}=c/m$. The modal velocity participation $C$ is irrelevant for computing ARMS Bandwidth of a monochromatic response, so we can compute $(\omega(t),\zeta(t))$ and then $\Delta\omega_{\text{ARMS}}(t)$, 
\begin{equation}
\label{sample:LTVsystem}
\Big(\omega(t),\zeta(t)\Big) = \Bigg(\sqrt{\frac{k+A\sin(t)}{m}},\frac{c}{2m\omega(t)} \Bigg).
\end{equation}
As shown in Fig. \ref{fig:TV} for varying strength of the time-varying stiffness coefficient, RMS Bandwidth grows in error due to the large variations in amplitude and therefore unsteadiness in the envelope leading to $\omega_{c} \neq 0$. While it is possible to obtain a more accurate RMS Bandwidth by tailoring enveloping parameters for each value of $A$, this is cumbersome and impractical for analyzing more than a handful of simulations. Conversely, ARMS Bandwidth provides a consistently accurate estimate for the modal damping coefficient. If the form of $\omega(t)$ is unknown, one can compute the Hilbert transform of the response, calculate the instantaneous frequency, and obtain an estimate for $\omega(t)$. For time-varying physical damping coefficient, such as $c(t)=c_{\text{static}}(1+\sin(t))$, ARMS Bandwidth provides a similarly accurate modal damping coefficient when $\omega(t)$ and $c(t)$ or $\zeta(t)$ are known.

\subsection{Guidelines for Numerical Computation of RMS Bandwidth}
While the RMS Bandwidth computation in Eq.~\eqref{sample:RMSBW} appears to be straightforward, there are important considerations regarding how the required FFT operation is performed. To this end, we consider a nonlinear oscillator with cubic hardening stiffness nonlinearity (referred to as the Duffing oscillator) similar to the one studied by Mojahed et al. and shown in Eq.~\eqref{sample:Duffing}, with small parameter $\epsilon=0.05$, damping coefficient $\lambda=0.25$, and time $t$ \cite{jsv2022alireza}: 
\begin{equation}
\label{sample:Duffing}
\ddot{x}(t) + \epsilon\lambda \dot{x}(t) + x(t) + \epsilon x^{3}(t) = 0.
\end{equation}
The system is given a relatively small initial velocity of 0.005 to ensure nonlinear effects are trivial and is integrated for $t\in[0,1000]$ with varying signal lengths $N_{\text{samples}}\in[2^{11},2^{15}]$. To determine the effect of signal length and FFT scheme on the RMS Bandwidth computation, the velocity envelope is fed to an N-point FFT code and an arbitrary-point FFT code. Furthermore, two versions of the envelope are generated, one with trailing zero-padding added prior to each FFT and the other without any modification. Note that the N-point FFT code will zero-pad any signal whose length is not a power of two while the arbitrary-point FFT code does not zero-pad.

\begin{figure}[H]
\centering
\includegraphics[width=1\textwidth]{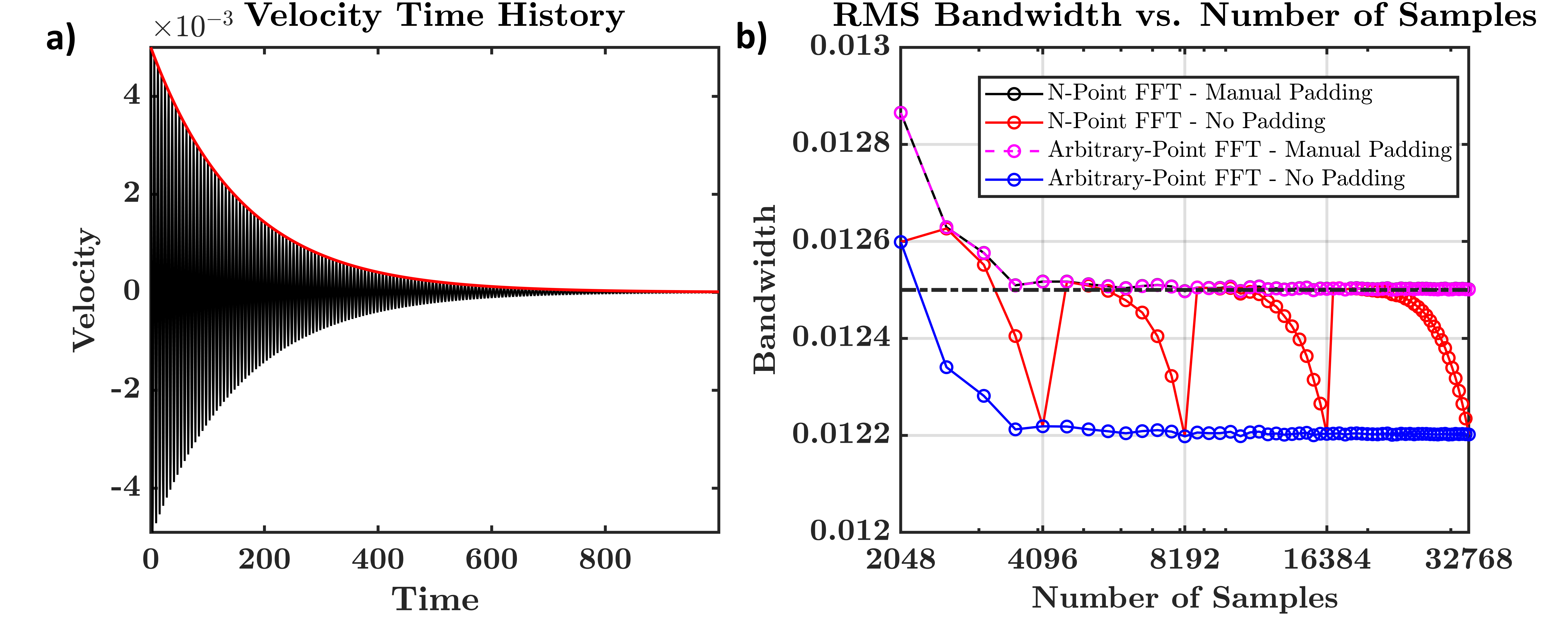}
\caption{Simulation results demonstrating the effect of data processing parameters: a) sample velocity time history and b) RMS Bandwidth effects; the linearized system modal damping coefficient is shown as the black dot-dashed line.}
\label{fig:FFTstudy}
\end{figure}

As observed with the linear SDOF system, a longer duration free decay greatly increases the accuracy of the RMS Bandwidth estimation; the same observation is made here when the signal is artificially lengthened via zero-padding prior to the FFT as shown in Fig. \ref{fig:FFTstudy}. The padded envelope RMS Bandwidth values converge to the expected modal damping coefficient. However, the unpadded envelope RMS Bandwidth values do not converge. The unpadded envelope arbitrary-point FFT bandwidth results in a consistent, but inaccurate value. The unpadded envelope N-point FFT bandwidth bounces between the unpadded arbitrary-point FFT results (lower bound) and both padded FFT results (upper bound). This is a direct result of the automatic padding performed by the N-point FFT routine to bring the number of samples to $2^{n}$ as required. When $N_{\text{samples}}=2^{n}$, the unpadded N-point FFT is equivalent to the unpadded arbitrary-point FFT. When $N_{\text{samples}}=(2^{n}+1)$, the unpadded N-point FFT is automatically padded with the greatest number of zeros to reach $2^{n+1}$ and therefore nearly coincides with the padded FFT. While the difference between the unpadded FFT RMS Bandwidth results and the exact solution is not particularly significant in this study, the difference increases drastically as simulation time is decreased. Exponential windowing to force the response to zero should not be used, as the RMS Bandwidth computed will be overestimated. All of these results are independent of the system natural frequency as long as the Nyquist criterion is satisfied. As discussed earlier, ARMS Bandwidth avoids such numerical complications related to simulation time and FFT operations. However, ARMS Bandwidth is not valid for nonlinear systems, unlike RMS Bandwidth. Ultimately, when calculating RMS Bandwidth, the signal should have a high signal-to-noise ratio and  decay to an extremely low final value with respect to the maximum value and/or be manually zero-padded prior to the FFT. In any case, the ARMS Bandwidth formula, Eq.~\eqref{sample:ARMSBWFINAL}, derived in this work alleviates these numerical inaccuracies altogether.

%-----------------------------------------------------------------
%% APPLICATION OF BANDWIDTH TO MDOF SYSTEMS---------------------------------
\section{Application of ARMS Bandwidth to MDOF Systems}\label{twoDOFsection}
\begin{figure}[H]
\centering
\includegraphics[width=0.4\textwidth]{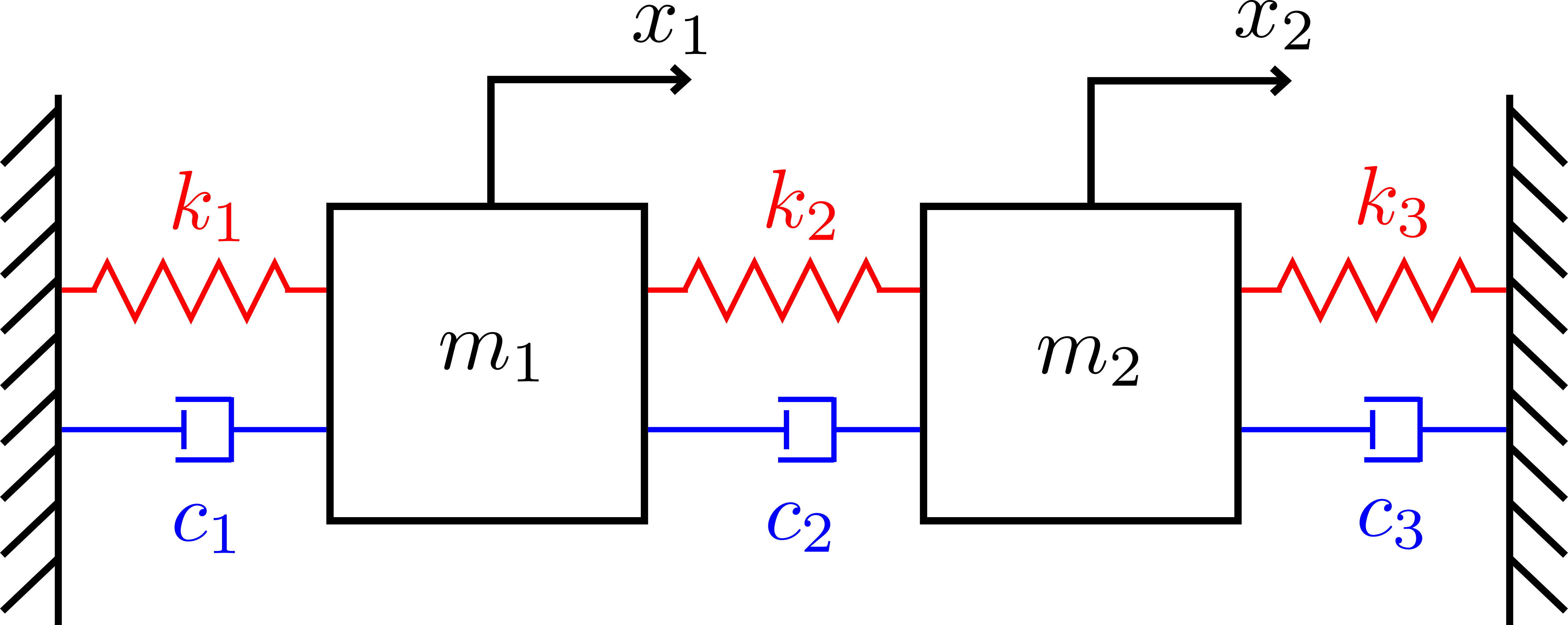}
\caption{Symmetric two-DOF system under investigation.}
\label{fig:2DOFsetup}
\end{figure}

We now validate the expression Eq.~\eqref{sample:ARMSBWFINAL} for the ARMS Bandwidth considering a numerical linear, time-invariant, classically damped MDOF system. Specifically, we investigate the bandwidth of the two-DOF system presented in Fig. \ref{fig:2DOFsetup} with system matrices shown in Eq.~\eqref{sample:MDOFeom},
\begin{equation}
\label{sample:MDOFeom}
\matr{M}=   
\begin{bmatrix}
    m_{1} &\! 0  \\ 
    0 &\! m_{2} 
  \end{bmatrix} 
  \;\;\;\; \matr{C} =   \begin{bmatrix}
    (c_{1}+c_{2}) &\! -c_{2} \\
    -c_{2} &\! (c_{2}+c_{3}) 
  \end{bmatrix}
  \;\;\; \matr{K} =   \begin{bmatrix}
    (k_{1}+k_{2}) &\! -k_{2} \\
    -k_{2} &\! (k_{2}+k_{3}) 
  \end{bmatrix}.
\end{equation}
The two masses are set to unity (again, units are omitted for convenience), the stiffnesses set to 200, and the damping coefficients set to 0.43333. These system parameters were chosen to produce pure in-phase and out-of-phase modes, each with a unique modal damping coefficient. In addition, the system is classically damped because $\matr{K}\matr{M}^{-1}\matr{C} = \matr{C}\matr{M}^{-1}\matr{K}$, i.e., the Caughey criterion is satisfied \cite{Caughey}. Solving the polynomial eigenvalue problem, $(\matr{M}\lambda^{2} + \matr{C}\lambda + \matr{K})\vect{\phi}=0$, we obtain an in-phase mode and out-of-phase mode with modal properties $(\omega_{1},\zeta_{1},\vect{\phi}_{1})=(14.1 \;\text{rad\slash s},0.015,[0.707 \;0.707]^{T})$ and $(\omega_{2},\zeta_{2},\vect{\phi}_{2})=(24.5 \;\text{rad\slash s},0.0265,[-0.707\;0.707]^{T})$, respectively, where $\omega_{i}$ denotes the $i^{th}$ natural frequency, $\zeta_{i}$ the $i^{th}$ modal viscous damping ratio, and $\vect{\phi}_{i}$ the $i^{th}$ mass-normalized modeshape. The system is integrated in time for 80 seconds. A total of 200 simulations are run, with each simulation generating random velocity initial conditions, $v_{1}(0),v_{2}(0) \in [-1,1]$, while keeping both initial displacements equal to zero. Equivalently, this provides a random sample of different modal kinetic energy distributions for bandwidth analysis. Note that when $v_{1}(0)=v_{2}(0)$, we excite purely the in-phase mode, and when $v_{1}(0)=-v_{2}(0)$, the out-of-phase mode. The RMS Bandwidth, Eq.~\eqref{sample:RMSBW}, is computed using the Akima-splined envelope of the sum of modal velocities, while the ARMS Bandwidth, Eq.~\eqref{sample:ARMSBWFINAL}, is computed using system natural frequencies, damping ratios, and modal velocity participations $C_{n}$ with $n=1,2,...,N$ based on Eq.~\eqref{sample:MDOFX}.

\begin{figure}[H]
\centering
\includegraphics[width=1\textwidth]{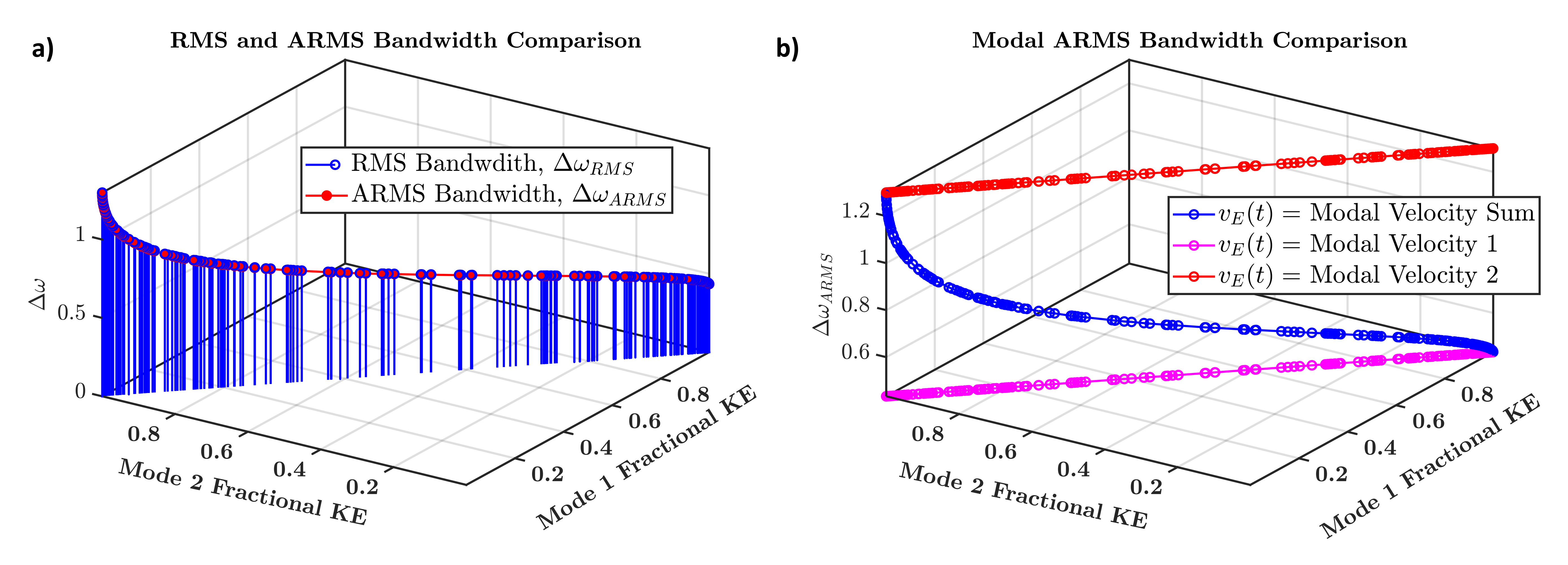}
\caption{Application of bandwidth to the two-DOF system for randomly varying initial velocities in the range $[-1,1]$ and zero initial amplitudes: a) system bandwidth comparison and b) ARMS Bandwidth for mode 1, mode 2, and the total multi-mode response. }
\label{fig:2DOFresults}
\end{figure}

The kinetic energy associated with each mode is computed, normalized by total modal kinetic energy, and plotted against both RMS and ARMS Bandwidth estimates as shown in Fig. \ref{fig:2DOFresults}. The RMS and ARMS Bandwidth values at each modal energy level agree well. The upper bandwidth limit, or maximum dissipative capacity of the entire two-DOF system, equals the second mode's modal damping coefficient of 1.3 and the lower bandwidth limit equals the first mode's modal damping coefficient of 0.43. These bandwidth limits occur when all of the system's kinetic energy is concentrated in a single mode, i.e., the bandwidth computed becomes the SDOF bandwidth for that mode, $\Delta\omega_{\text{ARMS}}=2\zeta_{n}\omega_{n}$. For multi-modal responses (i.e., when both the in-phase and out-of-phase modes participate in the free decay, but yet don't interact since they are uncoupled due to the classical damping distribution), the bandwidth is bounded, i.e., $(2\zeta_{n}\omega_{n})_\text{{min}} \leq \Delta\omega_{\text{ARMS}} \leq (2\zeta_{n}\omega_{n})_{\text{max}}$, and is a function of the distribution of modal energy, as expected.

%-----------------------------------------------------------------
%% APPLICATION OF BANDWIDTH TO CONTINUOUS SYSTEMS--------------------------
\section{Application of ARMS Bandwidth to Reduced-order Models of Continuous Systems}
It was demonstrated that the exact analytical expression for the ARMS Bandwidth, Eq.~\eqref{sample:ARMSBWFINAL}, agrees with the numerically evaluated integral expression for RMS Bandwidth for both SDOF and MDOF numerical models. We now demonstrate the applicability of these bandwidth expressions to a finite element ROM and the related physical model of an aircraft under impulsive excitation. This study aims to further demonstrate the broad applicability of ARMS Bandwidth to a broad class of practical engineering systems. Before we perform this study, however, it will be necessary to discuss a purely data-driven methodology that is developed specifically for computing ARMS Bandwidth, which extends its applicability to experimental modal test measurements. \emph{The aim is to enable computation of ARMS Bandwidth directly from measured time series without the need of any prior parametric modeling or related assumptions}.

\subsection{Data-driven Application of ARMS Bandwidth Methodology}\label{datadrivenmethod}
The expression for ARMS Bandwidth, Eq.~\eqref{sample:ARMSBWFINAL}, consists of SDOF-equivalent enveloped modal velocity participations, or equivalently, maximum modal amplitudes (measured immediately following the application of the impulse), natural frequencies, and damping ratios. Earlier application of ARMS Bandwidth to the two-DOF system relied on direct computation of the modal responses, which required the solution to the system's eigenvalue problem, $(\matr{M}\lambda^{2} + \matr{C}\lambda + \matr{K})\vect{\phi}=0 $. For practicing engineers, this solution is not always easily acquired since no explicit model of the tested structure necessarily exists to deliver the system matrices $\matr{M}$, $\matr{C}$, and $\matr{K}$. This underscores the need to develop an alternative purely data-driven methodology that eliminates the need to solve \emph{a priori} the eigenvalue problem, and instead relies solely on direct time series measurements. To this end, we now develop a purely data-driven methodology to extract the necessary quantities for the ARMS Bandwidth computation using the velocity time history at different points on a structure exclusively. In the case of acceleration measurements, the velocity time histories can be obtained by integrating high-pass filtered acceleration time histories collected during a modal test.

\begin{figure}[H]
\centering
\includegraphics[width=0.6\textwidth]{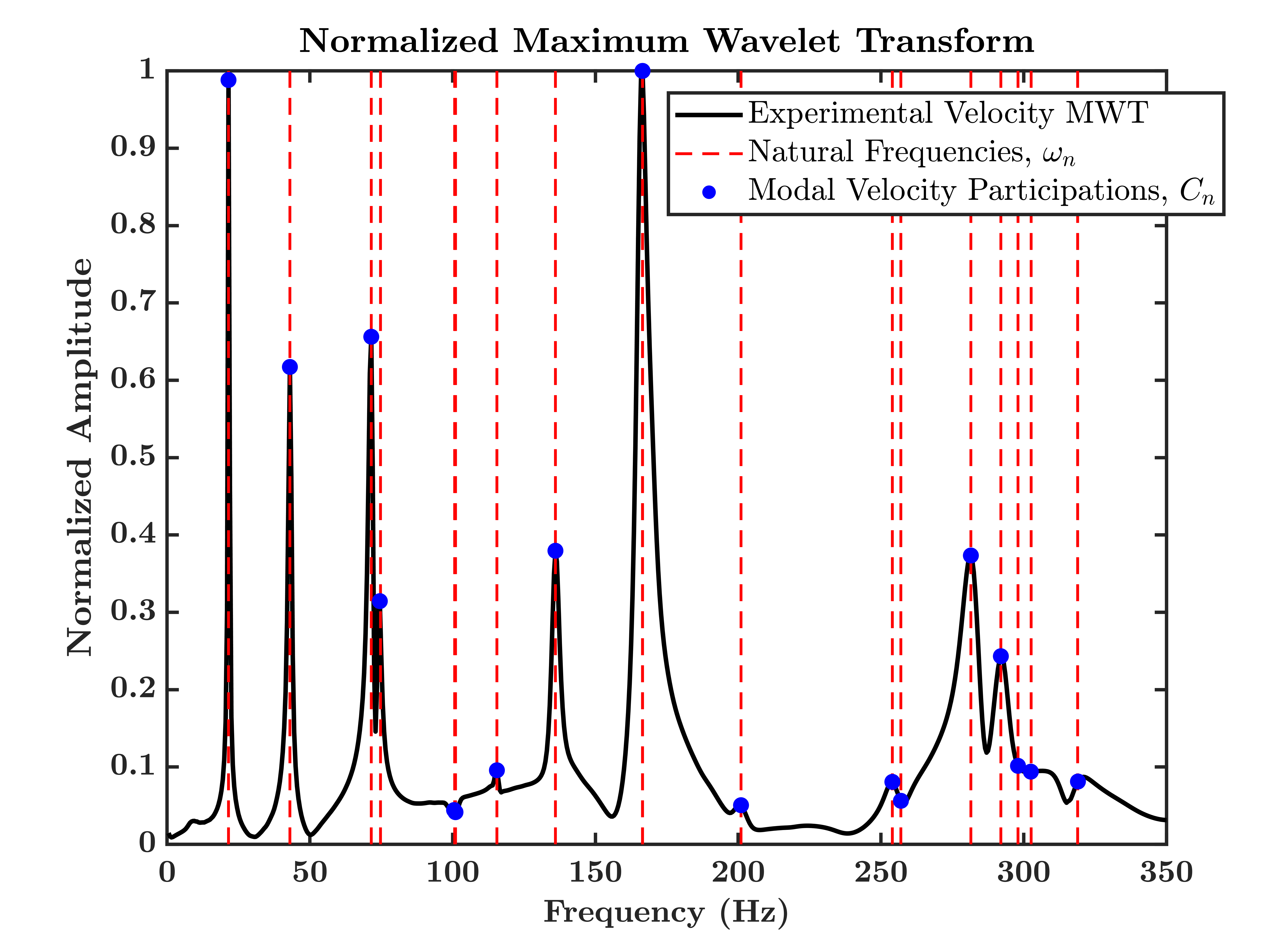}
\caption{Typical normalized (with the respect to the maximum peak) maximum wavelet transform plot of a linear MDOF system, showing the natural frequencies and extraction of modal velocity participations.}
\label{fig:mwtmethod}
\end{figure}

\begin{figure}[H]
\centering
\includegraphics[width=0.8\textwidth]{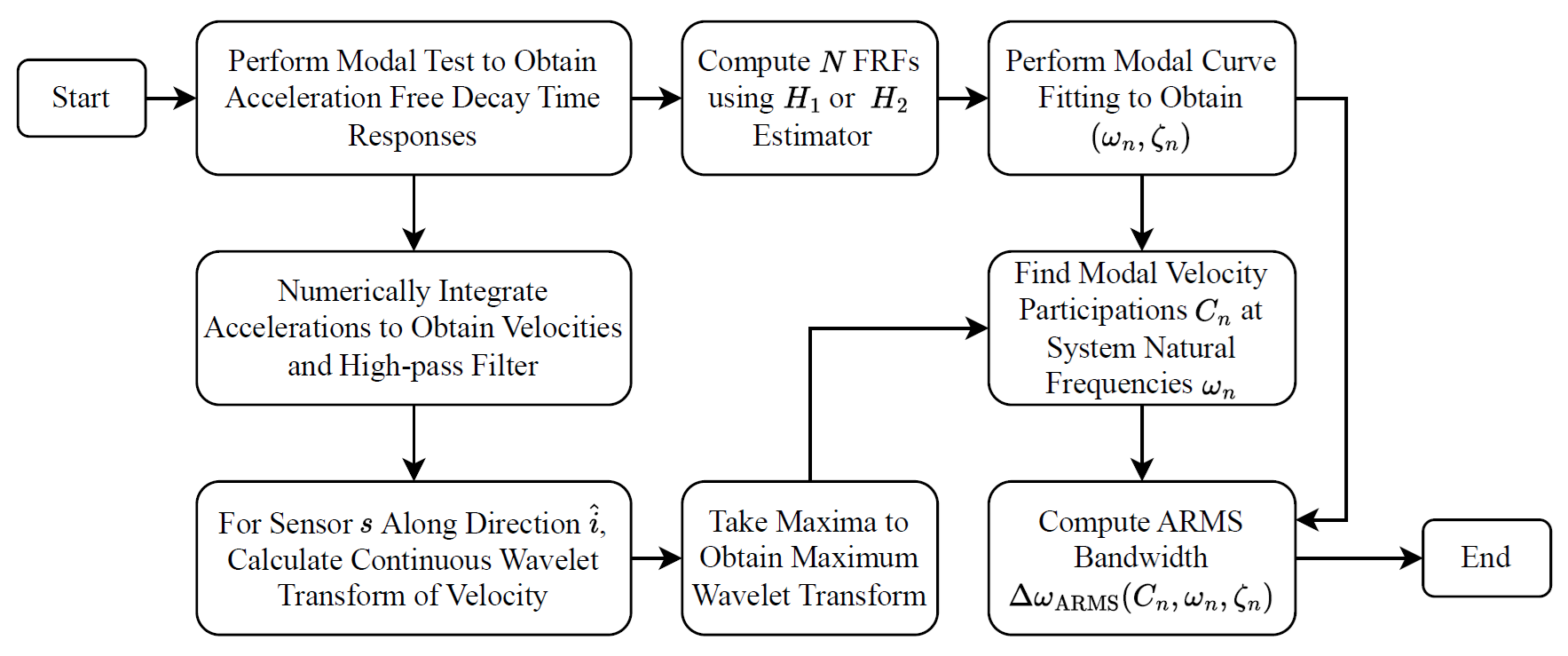}
\caption{Overview of purely data-driven methodology to compute ARMS Bandwidth from test data.}
\label{fig:flowchart}
\end{figure}

We assume a modal test, such as an impact hammer test, is performed and acceleration and force input time histories are collected simultaneously at a sufficiently high sample rate, which is typically at least $\frac{5}{4}(2F_{max})$ where $F_{max}$ is the maximum frequency of interest in Hertz and the $\frac{5}{4}$ factor accounts for anti-aliasing filter effects. Assuming $N$ frequency response functions (FRFs) are measured, one can perform modal curve-fitting to obtain estimates for the system's modal parameters $(\omega_{n},\zeta_{n},\vect{\phi}_{n})$ using standard modal analysis algorithms \cite{Ewins}. Separately, acceleration time histories can be integrated to obtain velocity time series, which can then be processed into the maximum wavelet transform using $L_{1}$ wavelet normalization via Eq. ~\eqref{sample:MaxWavelet}. The Appendix contains data and discussion to substantiate the use of $L_{1}$-normalization and provides additional modal participation extraction methodologies; all wavelet transform computations herein use $L_{1}$-normalization unless otherwise noted. Using the estimated natural frequencies, one can extract the modal participation $C_{n}$ of each modal velocity response from the maximum wavelet transform as shown in Fig. \ref{fig:mwtmethod} for a typical measurement of an MDOF system, then feed the estimated natural frequency, damping ratio, and modal participations into Eq.~\eqref{sample:ARMSBWFINAL} to compute the corresponding purely data-driven ARMS Bandwidth value. This methodology is summarized in Fig. \ref{fig:flowchart}. Hence, if the natural frequencies and damping ratios are already identified through modal analysis, this methodology can be applied to any free decay velocity time history of measured data. Furthermore, it was demonstrated in an auxiliary study provided in the Appendix that one may use the inverse wavelet transform or FRFs to extract the modal particpations needed in the ARMS Bandwidth computation.

\subsection{Physical Aircraft, Aircraft Finite Element Model, and Modal Test Data Descriptions}
\subsubsection{Physical Aircraft}
\begin{figure}[H]
\centering
\includegraphics[width=0.95\textwidth]{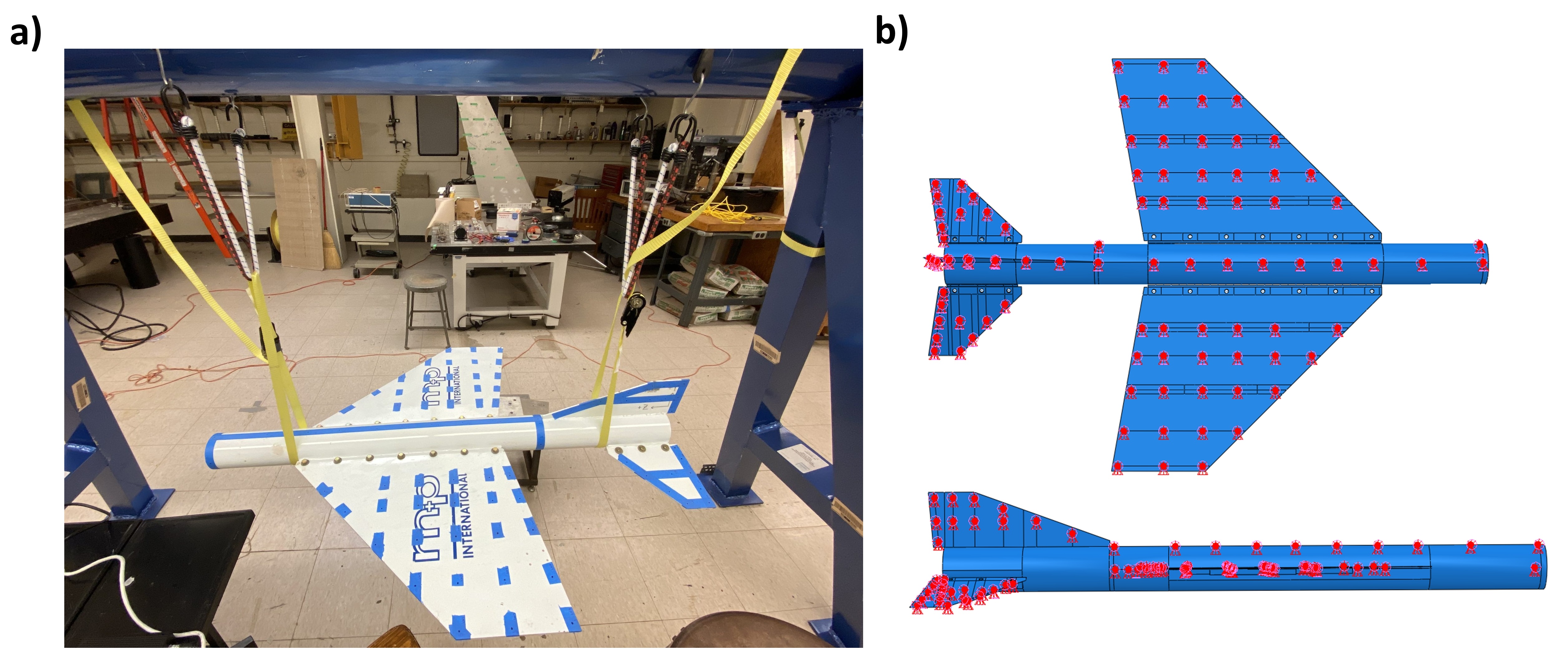}
\caption{Aircraft testbed: a) physical model suspended via shock cords and ratchet straps and b) Abaqus finite element model with retained nodes identified in red.}
\label{fig:AircraftFEM}
\end{figure}

The aircraft testbed is presented in Fig.~\ref{fig:AircraftFEM}a on the left. It is made out of steel and has a wingspan of 1.016 m, a nose-to-tail-tip length of 1.36 m, and a total mass of 29.54 kg. The horizontal stabilizer has an anhedral angle of approximately 30$^{\circ}$. The aircraft was suspended by looping ratchet straps around the body and connecting two shock cords to each strap to achieve free-free boundary conditions. The shock cords provided sufficient separation between rigid body and flexible body modes.

\subsubsection{Aircraft Finite Element Model}
The aircraft finite element model (FEM) is presented in Fig. \ref{fig:AircraftFEM}b on the right. The full-order FEM with 17,645 elements and 19,072 nodes is comprised of S4R shell elements with B31 beam connections representing bolted joints. A Craig-Chang ROM of the FEM was generated with 30 retained modes and 91 physical degrees-of-freedom spatially distributed across the aircraft. The retained nodes are shown in Fig. \ref{fig:AircraftFEM}b as red dots. The mass and stiffness matrices were extracted directly from the Abaqus normal modes and substructuring output file. The eigenvalue problem, $(\matr{M}_{\text{ROM}}\lambda^{2} + \matr{K}_{\text{ROM}})\vect{\phi}=0 $, was solved to obtain natural frequencies and the mass-orthonormalized modal matrix $(\vect{\omega},\vect{\Phi})$. Entries associated with rigid body modes or component synthesis modes were removed in the modal domain, i.e., those modal parameters ($\vect{\omega},\vect{\phi}$) were removed from the linear modal model of the aircraft. Modal damping ratios were identified based on curve-fitting of modal test data discussed in the following section. Simulations described in the subsequent sections were run by integrating the system in modal state-space form shown in Eq.~\eqref{sample:Statespace} using \emph{ode45} in MATLAB, 
\begin{equation}
\label{sample:Statespace}
  \dot{\vect{u}}=   
\begin{bmatrix}
    \matr{0} &\! \matr{I}  \\ 
    -\text{diag}(\omega^{2}) &\! -\text{diag}(2\zeta\omega)
  \end{bmatrix} 
  \vect{u} + \begin{bmatrix}
    \matr{0}  \\ 
    \matr{\Phi}^{T}\vect{F}
  \end{bmatrix}, 
\end{equation}where $\vect{u}=[\vect{\eta}\;\;\dot{\vect{\eta}}]^{T}$ is the modal state vector. Physical states were recovered using modal projection, $x=\matr{\Phi}\vect{\eta}$.

\subsubsection{Aircraft Modal Test Data}
\begin{figure}[H]
\centering
\includegraphics[width=0.6\textwidth]{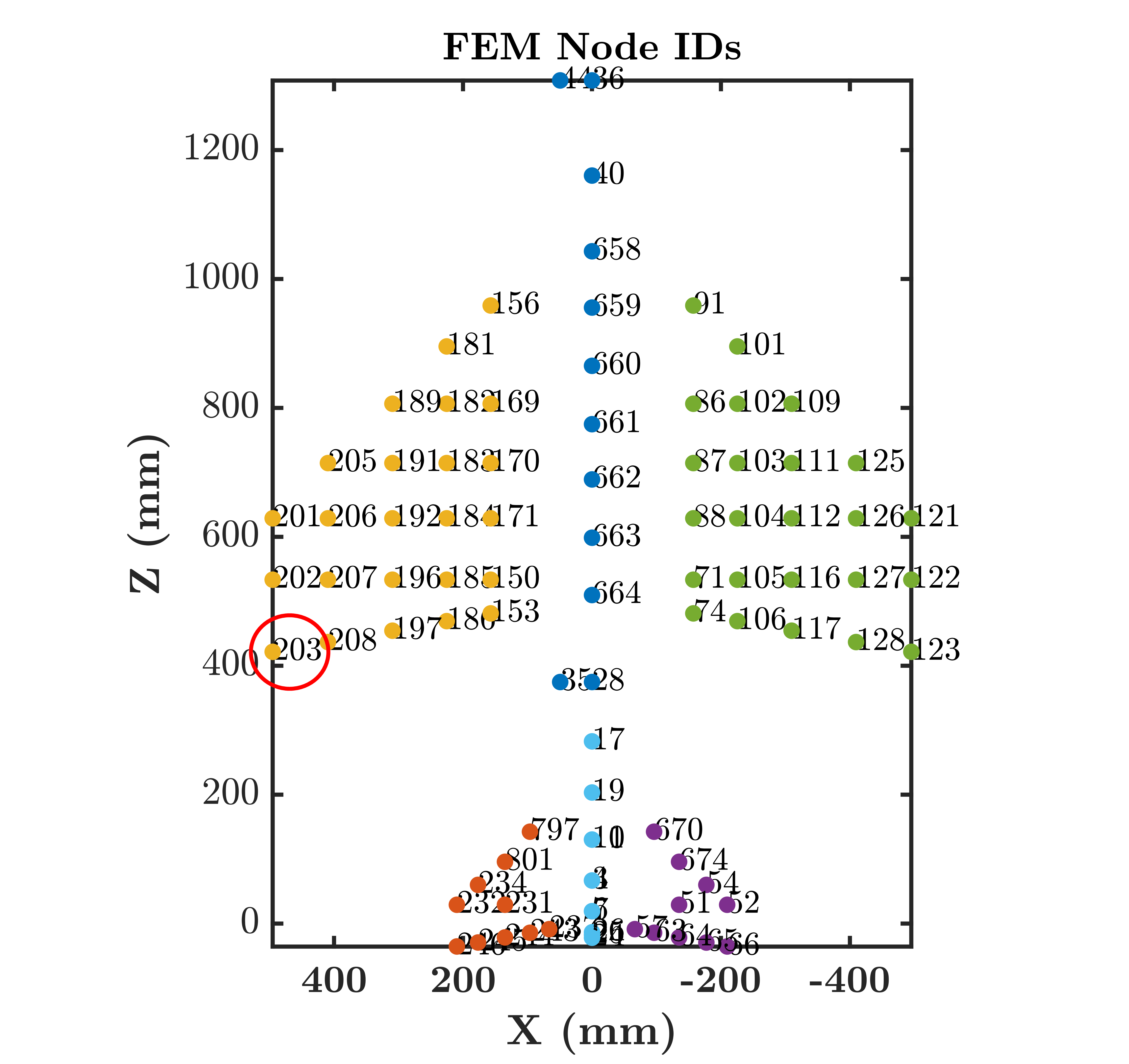}
\caption{Drive point at the aircraft port aft wingtip circled in red; all nodes in the figure correspond to retained physical degrees-of-freedom of the ROM.}
\label{fig:drivepoint}
\end{figure}

The physical aircraft was tested to collect baseline structural response data. A PCB Piezotronics Model 086C03 impact hammer along with a Model 356A45 triaxial accelerometer connected to a Crystal Instruments Spider-20HE were used to collect force input and acceleration output time series at a sample rate of 4,096 samples per second as throughput, i.e., no pre-processing other than anti-aliasing and digitization was performed. This sample rate satisfies the minimum sample rate of $\frac{5}{4}(2F_{max})$, as $F_{max}\approx 600$ Hz. While data at multiple locations across the aircraft were collected by roving the impact hammer, we focus on the drive point measurement shown in Fig. \ref{fig:drivepoint}.

\begin{figure}[H]
\centering
\includegraphics[width=1\textwidth]{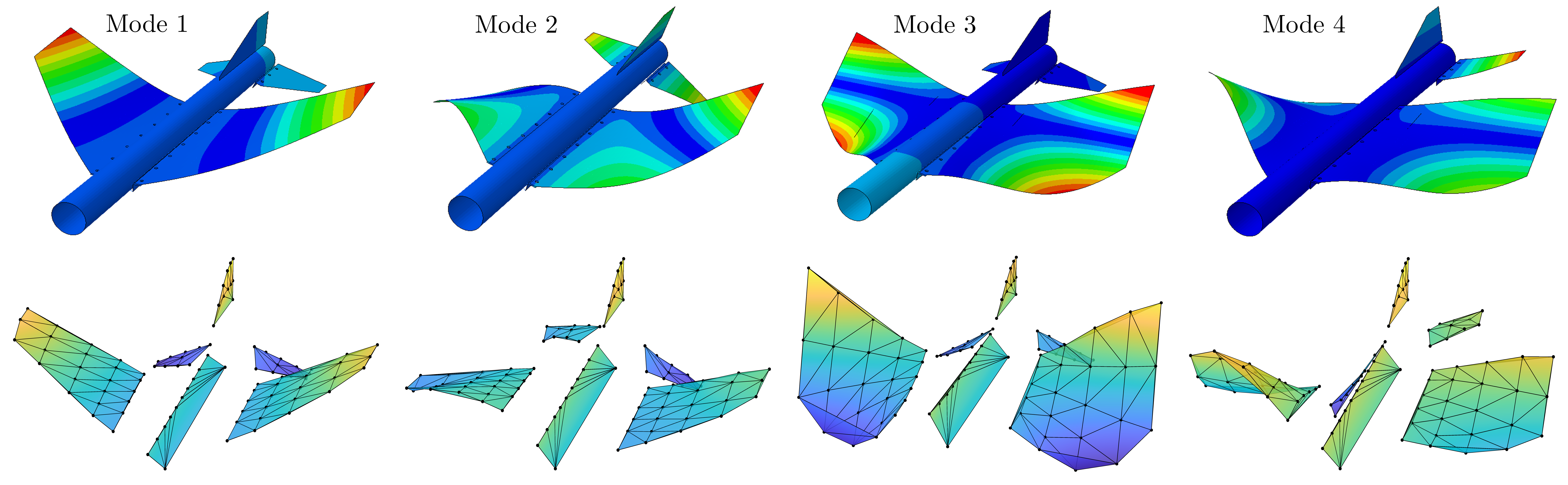}
\caption{First four modeshapes from aircraft ROM along the top row and associated estimated modeshapes from experimental modal testing along the bottom row.}
\label{fig:FEMmodeComparison}
\end{figure}

Least-squares complex exponential modal curve-fitting was performed in MATLAB to estimate the aircraft's modal parameters. A summary of the first four identified modes are provided in Table \ref{tab:modalID}. A comparison of the first four ROM and curve-fit modeshapes is provided in Fig. \ref{fig:FEMmodeComparison}; the associated natural frequencies of the curve-fit modal model are within 1.2\% of the ROM natural frequencies. The ROM was correlated to the physical aircraft response by tuning the initial estimate of modal damping ratios until the ROM response sufficiently reproduced the experimental response in the time, frequency, and time-frequency domains as discussed later in Section \ref{physrom}.

\begin{table}[htbp]
  \caption{Modal curve-fit results for first four modes.}
\makebox[\textwidth][c]{
    \begin{tabular}{ccc}

\cmidrule(lr){1-3}
$f\;(\text{Hz})$ & $\zeta\;(\%)$ & Modeshape Description \\ \midrule
 21.65 & 0.41 & 1\textsuperscript{st} Symmetric Vertical Wing Bending \\
 43.19 & 0.70 & 2\textsuperscript{nd} Antisymmetric Vertical Wing Bending \\
 71.82 & 0.50 & 1\textsuperscript{st} Symmetric Wing Torsion \\
 74.09 & 0.48 & 1\textsuperscript{st} Antisymmetric Wing Torsion \\
\bottomrule
\end{tabular}}
\label{tab:modalID}%
\end{table}%

\subsection{Finite Element Model Aircraft Bandwidth}
We study the bandwidth of the aircraft ROM under impulsive excitation. We first investigate drive point bandwidth from constant-energy pulses of varying duration to demonstrate the connection between initial modal energy distribution and bandwidth. We then investigate the spatial distribution of bandwidth resulting from a single pulse to demonstrate the efficacy of ARMS Bandwidth to quantify the dissipative capacity of a structure at different forcing locations along an unchanging forcing direction. To the authors' best knowledge, \emph{the ARMS Bandwidth presented in this work provides for the first time in the engineering literature a single scalar quantitative measure to accurately predict the overall dissipative capacity of an MDOF oscillating structure subjected to an applied impulse at a certain location along a certain direction}. Moreover, what makes the ARMS Bandwidth especially suitable to engineering practice and design, is that expression Eq.~\eqref{sample:ARMSBWFINAL} is directly related to the modal parameters and the modal participations of the structural modes that comprise the measured structural response, so the ARMS Bandwidth estimation can be performed in tandem with traditional modal testing measurements, e.g., with modal hammer impact tests.

\subsubsection{Local ARMS Bandwidth from constant-energy but variable-duration Pulses}
Constant input energy pulses of varying duration are applied to the ROM in Eq.~\eqref{sample:Statespace} at the port aft wingtip (see Fig. \ref{fig:drivepoint}) in the vertical direction to study the effect of pulse duration on bandwidth. The equation for a half-sine pulse as a function of signal energy $E_{\text{p}}$  and pulse duration $t_{d}$ is shown in Eq.~\eqref{sample:Impulse}, 
\begin{equation}
\label{sample:Impulse}
f(t)= \sqrt{\frac{2E_{\text{p}}}{t_{d}}}\sin\Big(\frac{\pi}{t_{d}}t \Big), \; 0\leq t \leq t_{d}.
\end{equation}
\begin{figure}[H]
\centering
\includegraphics[width=1\textwidth]{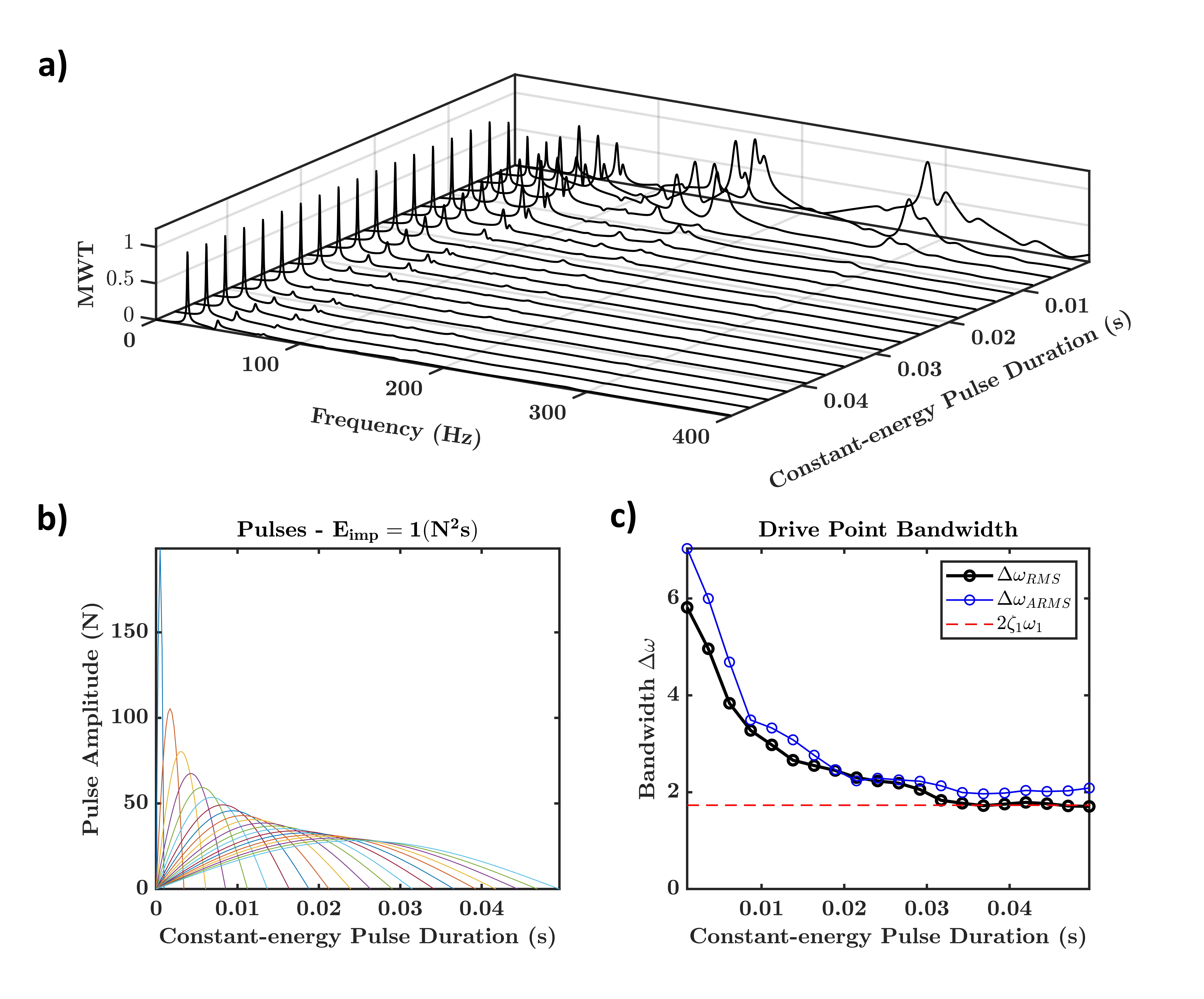}
\caption{Drive point bandwidth study for the ROM in Eq.~\eqref{sample:Statespace}: a) maximum wavelet transform waterfall plot demonstrating the relationship between pulse duration and initial modal excitation, b) constant-energy pulses applied to the ROM, and c) comparison of drive point RMS and ARMS Bandwidths.}
\label{fig:ConstantE}
\end{figure}
The local RMS Bandwidth is computed numerically as described previously using the drive point velocity envelope. For comparison, the local ARMS Bandwidth is also computed using the data-driven methodology with the drive point velocity as input. The results of this study are summarized graphically in Fig. \ref{fig:ConstantE}. Increasing the pulse duration results in a modal energy distribution that favors excitation of mainly low frequency modes, resulting in a lower bandwidth and thus lower dissipative capacity. This is related to the fact that the bandwidth definitions Eq.~\eqref{sample:RMSBW} and Eq.~\eqref{sample:ARMSBWFINAL} are related to the variance of the energy distribution in the frequency domain of the meassured response. As such, it is natural that bandwidth is directly related to the frequency response of the structure, as indicated by the maximum wavelet transforms presented in Fig. \ref{fig:ConstantE}. Decreasing pulse duration excites a broader range of modes so it expands the number of structural modes participating in the frequency response; in turn, this increases the bandwidth since high frequency modes tend to have a greater modal damping coefficient and thus dissipate energy more efficiently, at a faster rate. In the limit of very small pulse duration, $t_{d}\rightarrow 0$, the pulse approaches the Dirac delta function, $\lim_{t_{d} \to 0} f(t) = \delta(t)$, whose Fourier transform is unity for all frequencies $F(\omega)=1$ resulting in equal excitation of all modes (theoretically). Conversely, for $\lim_{t_{d}\to K} f(t)$ where $K$ is an arbitrary non-zero positive number, the Fourier transform of the pulse resembles a low-pass filter, resulting in sufficient excitation of modes at frequencies below the cutoff frequency. 

Shown in the bottom right is a direct comparison of bandwidth values for the drive point. The RMS and ARMS Bandwidth agree reasonably well for each pulse considering the variation introduced to the former in the enveloping process. However, \emph{ARMS Bandwidth more accurately captures the modal contributions to bandwidth}. As the pulse approaches its longest duration, the modes excited tend to the first and second modes, with the first mode being significantly more dominant than the second as shown in the left-most bottom curve in the top of Fig. \ref{fig:ConstantE}. Note, however, that for the same limit of relatively long pulses, the corresponding limits for the two bandwidth measures approach different limits, namely $\Delta\omega_{\text{RMS}}\rightarrow 2\zeta_{1}\omega_{1}$ and $\Delta\omega_{\text{ARMS}}\rightarrow 2\zeta_{1}\omega_{1} + f(2\zeta_{2}\omega_{2})$, where $f$ is some function that captures the contribution of the second mode. Therefore, we conclude that ARMS Bandwidth captures the added dissipative capacity associated with the low-amplitude second mode, whereas RMS Bandwidth does not. This demonstrates that the enveloping process and use of the FFT of the envelope associated with RMS Bandwidth can result in low-amplitude modes hidden by dominant modes, therefore producing an underestimated bandwidth. This is also true in a general sense; RMS Bandwidth tends to underpredict the true dissipative capacity of the system that \emph{is} accurately captured by ARMS Bandwidth. Additional studies with the same range of pulse durations with different energy levels demonstrate that $\frac{d}{dE_{\text{p}}}\Big(\Delta\omega_{\text{ARMS,RMS}}\Big)=0$ for any given $t_{d}$, i.e., that both bandwidth estimates are independent of the input energy. This is a general result; bandwidth does not change with pulse energy assuming the pulse form and duration do not change, as the relative modal energy distribution and system modal parameters are independent of input energy level. This result is consistent with the physics of the problem at hand; indeed, due to the linearity in the measured dynamics, the results are expected to be independent of energy (this is not the case, however, when nonlinearities affect the measured response).

\subsubsection{Spatial Distribution of Local ARMS Bandwidth from a Single Pulse}
To determine the spatial distribution of bandwidth due to a vertical pulse of the form of Eq.~\eqref{sample:Impulse} with $(E_{\text{p}},t_{d})=(1.8,0.004\;\text{s})$ at the port aft wingtip (see Fig. \ref{fig:drivepoint}), the ARMS Bandwidth is computed based on the resulting velocity at each structural node (or retained DOFs) of the ROM in the vertical direction (denoted by the unit vector $\hat{i}_{3}$) using the data-driven methodology previously discussed. Moreover, we compute the \emph{relative} ARMS Bandwidth by subtracting the mean ARMS Bandwidth across the aircraft from the local ARMS Bandwidth as shown in Eq.~\eqref{sample:relativebw},
\begin{equation}
\label{sample:relativebw}
\Big[ \Delta\omega_{\text{ARMS}}(\vect{r}_{k},\hat{i}_{3}) \Big]_{\text{rel.}} =  \Delta\omega_{\text{ARMS}}(\vect{r}_{k},\hat{i}_{3}) - \frac{1}{N}\displaystyle\sum_{k=1}^{N} \Big( \Delta\omega_{\text{ARMS}}(\vect{r}_{k},\hat{i}_{3}) \Big).
\end{equation}
\begin{figure}[H]
\centering
\includegraphics[width=1\textwidth]{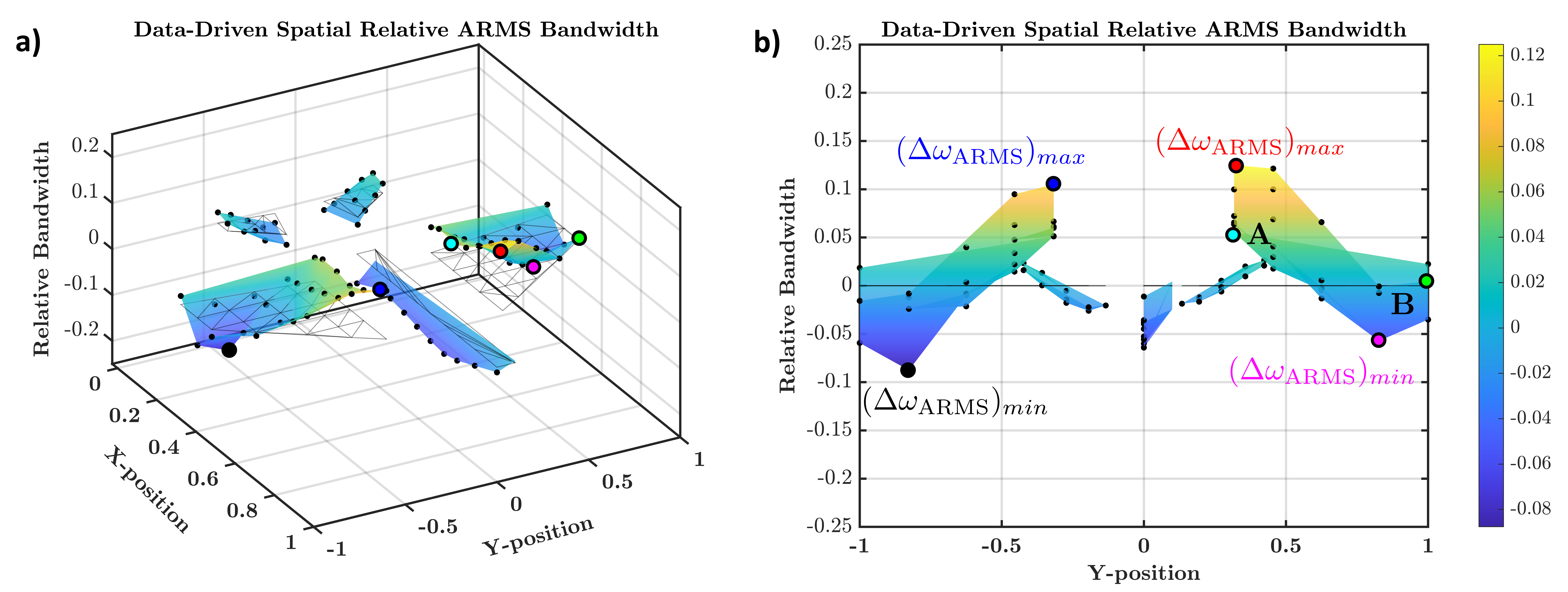}
\caption{Relative ARMS Bandwidth across the aircraft due to impulsive vertical excitation at the port aft wingtip based on the ROM: a) isometric view and b) front view with minimum, maximum, and slightly above and near average (locations A and B, respectively) wing relative ARMS Bandwidth locations identified.}
\label{fig:bwmap}
\end{figure}In this computation, the mean local ARMS Bandwidth is taken across the entire aircraft; however, for a full-scale aircraft, it may be more useful to present relative bandwidth for each lifting surface. Accordingly, the spatial distribution of ARMS Bandwidth here is computed at the drive point and all other (non-drive) points using a single simulation with force input at the single drive point. Alternatively, one could get the same spatial distribution of ARMS Bandwidth using only drive point responses, i.e., by recording the vertical component of the velocity at each location where the pulse is applied; this is discussed in the Appendix.

\begin{figure}[H]
\centering
\includegraphics[width=0.6\textwidth]{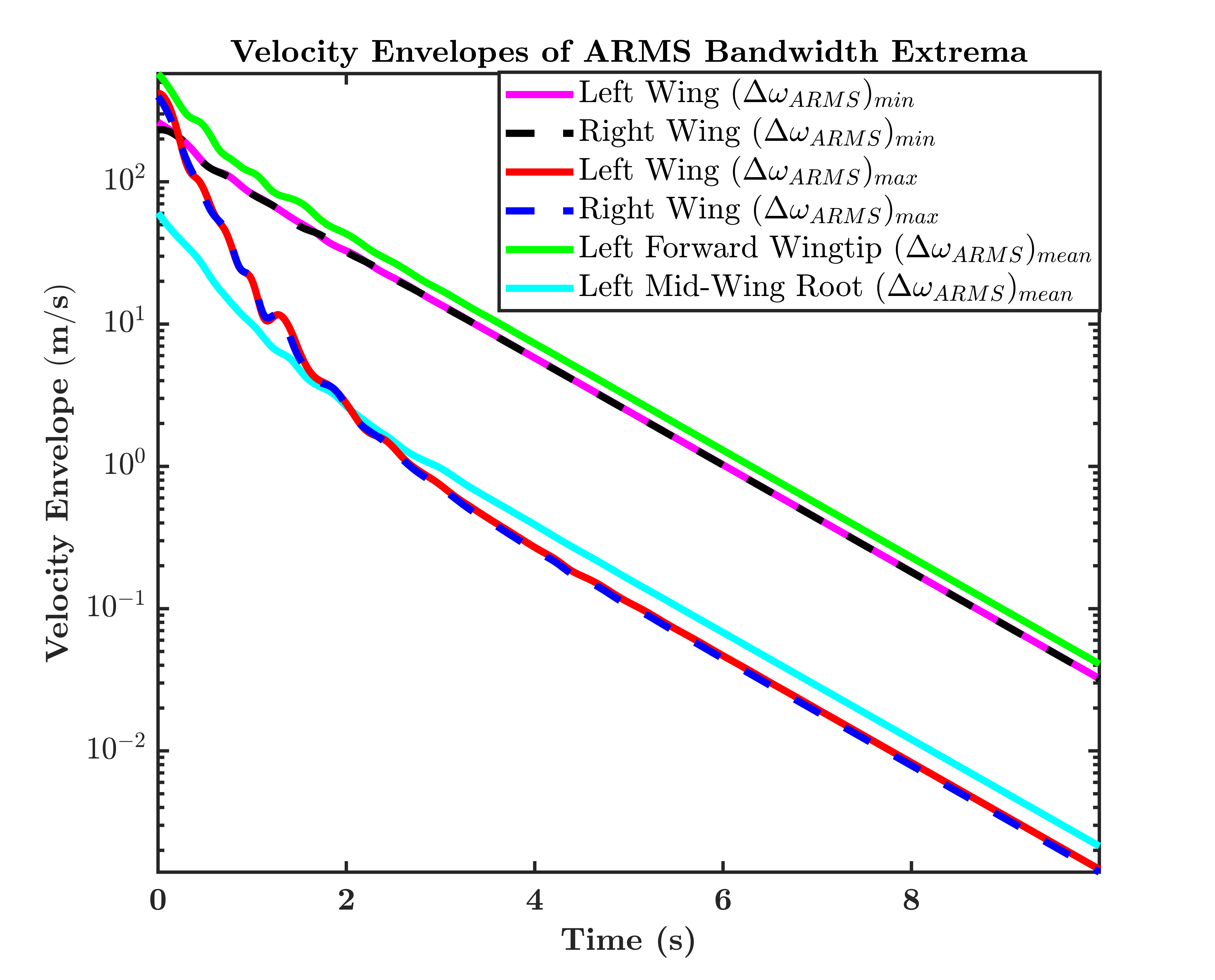}
\caption{Velocity envelopes at points of maximum and minimum $\Delta\omega_{\text{ARMS}}$ in addition to two interior points, A and B, slightly above and below average $\Delta\omega_{\text{ARMS}}$, respectively, based on the ROM; these results directly demonstrate that spatial bandwidth accurately quantifies the spatial dissipative capacity of the structure for a given impulsive loading at a given location and direction.}
\label{fig:minmaxenvelope}
\end{figure}

Spatial bandwidth results are presented in Fig. \ref{fig:bwmap}. There are clear regions of higher relative bandwidth (greater dissipative capacity) and lower relative bandwith (lower dissipative capacity) of the aircraft model. Specifically, relative bandwidth tends to increase as we approach the wing root and decrease as we approach the wingtips. This intuitively makes sense, as fundamental modes such as the first wing vertical bending and first torsional modes have maximum modal displacement toward the wingtip. The maximum bandwidth occurs toward the inner leading edge of the wing and the minimum toward the wingtip mid-chord as highlighted in Fig. \ref{fig:bwmap} on the right. The dissipative capacity at these locations is visualized in Fig. \ref{fig:minmaxenvelope}. The velocity envelopes at the minimum bandwidth locations decay mostly at the same rate (constant slope) associated with modal participation of fundamental modes. The velocity envelopes at the maximum bandwidth locations initially decay very quickly (steep slope) due to modal participation of higher-order modes with greater modal damping coefficients. Once those high-order modes decay, the envelope decay rate (slope) tends to that of the minimum bandwidth location, as only the lightly-damped fundamental modes remain. Also shown is the envelope at the left forward wingtip (point B), which has a relative bandwidth near the mean bandwidth, and the envelope at the left mid-wing root (point A), which has a relative bandwidth slightly above the mean bandwidth. As expected, the envelope at point B has a steeper initial slope than the minimum bandwidth envelope while the envelope at point A is initially steeper compared to point A and the minimum bandwidth locations. These results demonstrate that one can accurately visualize (and quantify) the spatial dissipative capacity of a structure by computing the spatial distribution of relative ARMS Bandwidth, a finding which has applications for structural modification and vibration mitigation (e.g., placement of tuned mass dampers), among many others.

\subsection{Physical and ROM ARMS Bandwidth from an Experimental Pulse}\label{physrom}
\begin{figure}[H]
\centering
\includegraphics[width=1\textwidth]{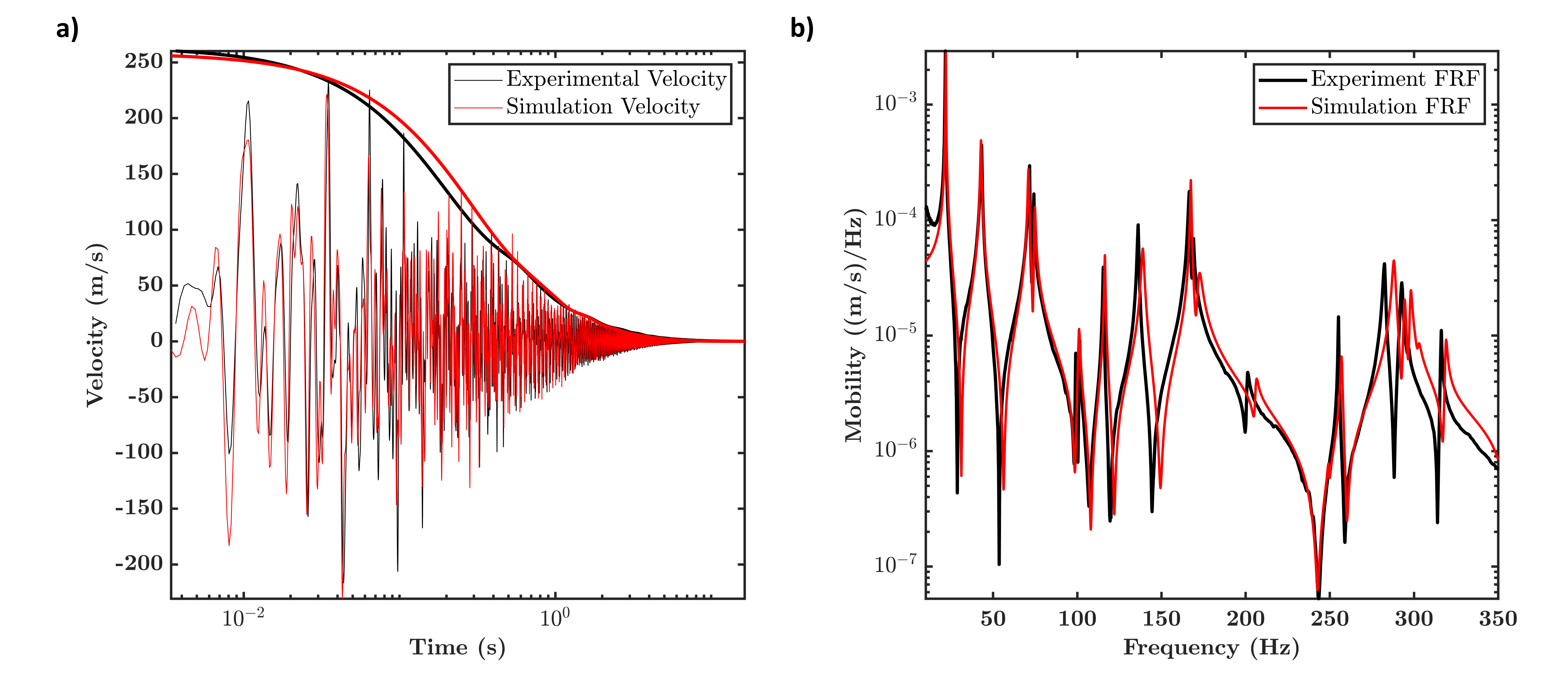}
\caption{Comparison of experimental and simulation drive point responses: a) velocity time histories and envelopes and b) and mobility frequency response functions.}
\label{fig:velocityhistorycomp}
\end{figure}

\begin{figure}[H]
\centering
\includegraphics[width=1\textwidth]{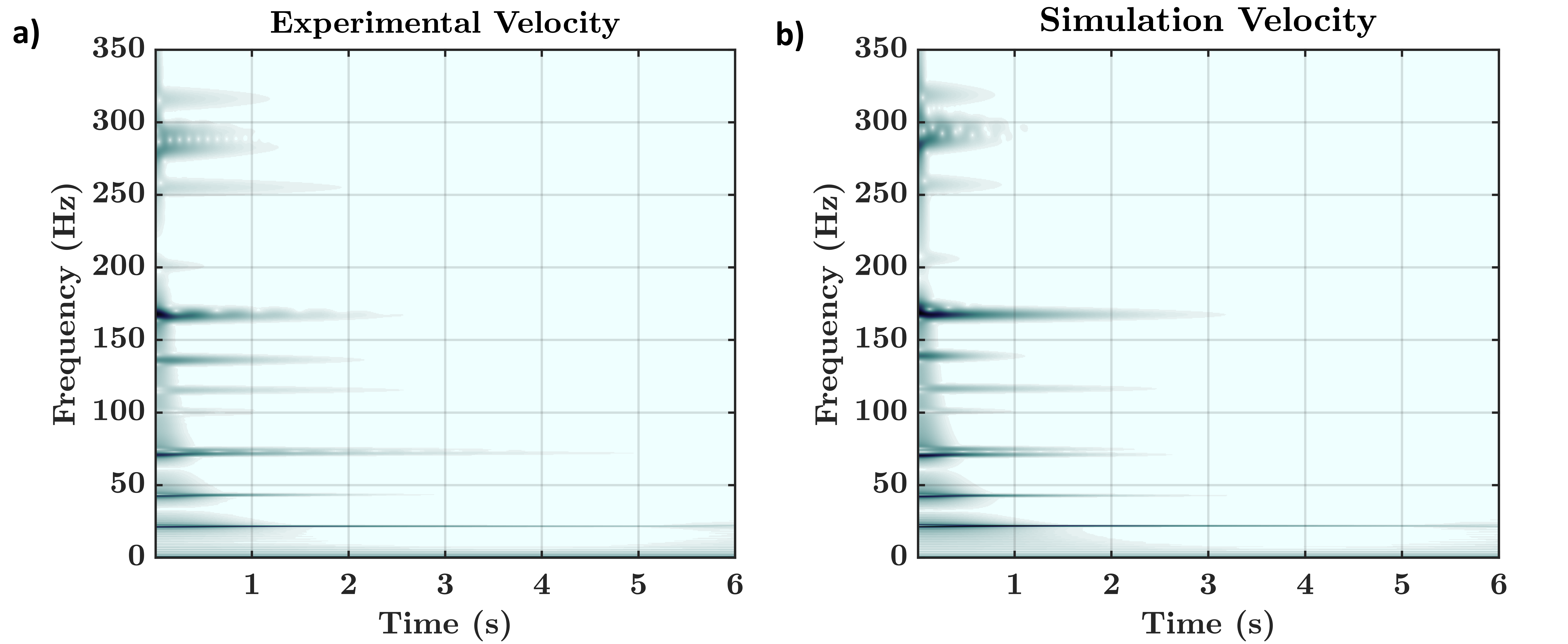}
\caption{Comparison of experimental and simulation time-frequency drive point responses over the first 6 seconds from the same experimental pulse: wavelet transform of the drive point a) experimental velocity and b) ROM simulation velocity.}
\label{fig:velocitywaveletcomp}
\end{figure}

Bandwidth computations for the aircraft model until now have been applied to the numerical ROM in Eq.~\eqref{sample:Statespace} only. Taking the bandwidth computation one step further, structural response data from an impact hammer test of the physical aircraft model was collected as discussed earlier, so we can apply the experimental pulse to the ROM and compute the local bandwidth for both the ROM and physical aircraft at the same drive point. For the ROM simulation, the experimental pulse is linearly interpolated for $t\leq t_{d}$, where $t_{d}=0.003418$ second. In addition, the simulation duration matches the experimental response duration of 16 seconds. For the physical aircraft model, the drive point acceleration due to the pulse was measured during modal testing and was therefore immediately available for the bandwidth computation. The experimental drive point acceleration was passed through a zero-phase high-pass finite impulse response filter, numerically integrated, then filtered again. To demonstrate dynamic similitude between the ROM and physical aircraft responses, a direct comparison of the experimental and correlated ROM simulation drive point velocity time histories and their corresponding wavelet transforms are provided in Figs. \ref{fig:velocityhistorycomp} and \ref{fig:velocitywaveletcomp}, respectively. Very good agreement between the experimental and ROM results is noted, validating the efficacy of our numerical predictions.


Both local RMS and ARMS Bandwidths are computed as described previously based on the drive point velocity free decays and are presented in Table \ref{tab:aircraftBW_table}. The numerically-exact simulation ARMS Bandwidth is computed using $C_{in}=\text{max}(|\phi_{in}\dot{\eta}_{n}|)$, where $i$ is the measurement DOF and $n$ is the mode number. The RMS Bandwidths for experimental and simulation data are within approximately 2\% of one another, as are the ARMS Bandwidths. Furthermore, using simulation data, the numerically-exact ARMS Bandwidth is within approximately 0.2\% of the data-driven ARMS Bandwidth. However, the ARMS and RMS Bandwidths differ significantly, with RMS Bandwidths being underpredicting, since the enveloping process associated with RMS Bandwidth tends to subsume less dominant frequency content and thus trivializes their dissipative capacity contribution. Differences between simulation and experimental bandwidths for either RMS or ARMS Bandwidth are attributed to the natural frequencies and damping ratios used in ARMS Bandwidth being estimates and the fact that the experimental aircraft model has joints that could introduce unmodelled physics, such as non-classical (non-proportional) damping distribution and friction. We note at this point that the ROM includes modes up to 550 Hz and the experimental velocity FRF and wavelet transform show the dominant responses at frequencies below 612 Hz. As a result, the simulation and experimental ARMS or RMS Bandwidths are very similar. However, if a shorter duration experimental pulse, e.g., using an impact hammer with a steel tip, was applied to both the physical aircraft and ROM, assuming the ROM contains the same modes up to 550 Hz, we would likely obtain a much greater bandwidth using experimental data compared to simulation data given that the ROM does not model those high frequency modes and therefore does not capture the dissipative capacity associated with them. This study demonstrates the accuracy and applicability of ARMS Bandwidth to both modal test correlated (dynamically-representative) structural ROMs and modal test data from a continuous, physical structure and, in this case, an experimental model aircraft. 

\begin{table}[htbp]
  \caption{Comparison of aircraft drive point local bandwidth.}
\makebox[\textwidth][c]{
    \begin{tabular}{cccccc}
\toprule
\multicolumn{3}{c}{Simulation} & \multicolumn{2}{c}{Experiment} \\
\cmidrule(lr){1-3}\cmidrule(lr){4-5}
$\Delta\omega_{\text{RMS}}$  & $\Delta\omega_{\text{ARMS}}$ & $\Delta\omega_{\text{EXACT ARMS}}$ & $\Delta\omega_{\text{RMS}}$  & $\Delta\omega_{\text{ARMS}}$ \\ \midrule
 3.59 & 5.30 & 5.34 & 3.55 & 5.23 \\
\bottomrule
\end{tabular}}
\label{tab:aircraftBW_table}%
\end{table}%

%% CONCLUSIONS----------
\section{Conclusions}
In this work, we established and validated ARMS Bandwidth, which quantifies the dissipative capacity of an MDOF, linear, time-invariant system with classical viscous damping distribution. The ARMS Bandwidth in expression Eq.~\eqref{sample:ARMSBWFINAL} developed in this work enhances the usefulness and applicability of the current RMS Bandwidth in expression Eq.~\eqref{sample:RMSBW}, as it eliminates the need for envelope fitting, which for multi-modal responses can be a challenging task and a source of numerical inaccuracies. In fact, to our best knowledge, this work is the first to extend the classical concept of bandwidth (defined for SDOF linear, lightly damped oscillators) to a general, linear classically viscously damped MDOF dynamical system. The efficacy of this concept was demonstrated by applying and comparing the ARMS and RMS Bandwidth expressions to SDOF systems, MDOF systems, a ROM aircraft system, and a physical (experimental) system to characterize their dissipative capacity, describe the underlying dynamics of free decay, and generate guidelines for computing bandwidth in practical applications. Moreover, the purely data-driven methodology for computing ARMS Bandwidth that is developed herein extends its applicability to experimental data, and, for the first time enables an engineer to assess the overall dissipative capacity at a given location and direction of a practical structure (even with a complex configuration). In addition, one can now assess the spatial distribution of the ARMS Bandwidth over the entire structure, purely through modal testing measurements. In that regard, one could consider the ARMS Bandwidth as a very useful extension of the classical modal impact hammer test that is used extensively in current modal testing. A summary of the main findings of our study regarding characteristics of bandwidth computation is provided below.

In general, the dissipative capacity of a system quantifies its effectiveness at dissipating energy over time. In that context, RMS and ARMS Bandwidth quantify the overall dissipative capacity of a structure, being an SDOF linear oscillator or single structural mode, or an MDOF ROM of a linear structure with classical damping. We note at this point that the RMS Bandwidth is still applicable to MDOF ROMs with non-classical damping, nonlinearities or time-varying system parameters; however, in that case envelope fitting is required, and no analogous analytical expression to ARMS Bandwidth is available yet to relate it to system parameters (although current efforts address extending ARMS Bandwidth to non-proportionally damped MDOF systems).

As a check, it was shown that for a linear, time-invariant SDOF oscillator (or single structural mode), the ARMS Bandwidth recovers the modal damping coefficient, $\Delta\omega_{\text{ARMS}}=2\zeta\omega$, as expected. For MDOF linear systems, the bandwidth depends on the initial modal excitations, so its minimum or maximum values converge to the modal damping coefficients of the lightest and heaviest damped modes, respectively. It follows, that for continuous (practical) structures, the maximum bandwidth is bounded by the maximum modal damping coefficient of the modes that are excited by an impulse excitation (e.g., an impact hammer test).

In synopsis, the bandwidth of an MDOF system or ROM is a function of initial modal energy distribution and system modal parameters, and thus, is tied by the initial excitation of its modes following the application of an impulse load; this follows from the fact that the bandwidth is computed exclusively from the free decay of the structural dynamics, and is an inherent property of the system. Moreover, it was shown that, due to linearity, bandwidth is energy independent for a given pulse form and duration. The ARMS Bandwidth expression Eq.~\eqref{sample:ARMSBWFINAL} derived in this work avoids error-prone and manual definitions of enveloping associated with RMS Bandwidth, and can be computed using the purely data-driven methodology developed herein.

The analytical expressions and data-driven methodologies developed herein are valid for and applicable to a broad class of linear MDOF and ROMs with classical damping distribution. It is of great interest and importance to extend this work to systems with non-classical damping, as complex and closely-spaced modes introduce energy exchanges between vibration modes and, therefore, new paths for energy dissipation. In addition, it is important to extend this work to weakly and strongly nonlinear systems, as nonlinear resonances and energy exchanges through harmonic generation and multi-frequency energy transfers greatly impact the dissipative capacity of such systems. While the classical time-bandwidth limit was not discussed here, the ARMS Bandwidth expression directly relates a system's modal parameters to the dissipative capacity of the system, and therefore the typical time-bandwidth limit for linear time-invariant SDOF systems of $\Delta t \Delta \omega\approx 1$, where $\Delta t$ is characteristic time, or decay time constant, and $\Delta \omega$ is bandwidth, will be investigated in a subsequent work \cite{gaborCommunication}. In subsequent work, we aim to address direct application of bandwidth to practical problems in aerospace systems, such as vibration mitigation, energy harvesting, and flight test data processing and monitoring.

\section*{Appendix}\label{appndx}
\subsection{SDOF ARMS Bandwidth Derivation} 
We start by computing the Fourier transform of the assumed velocity time series, Eq.~\eqref{sample:SDOFX}, and take its modulus to the fourth power,
\begin{equation}
V_{E}(\omega) = C_{n} \displaystyle\int_{-\infty}^{\infty} e^{-(\zeta_{n}\omega_{n} + j\omega)t} dt = C_{n} \frac{\zeta_{n}\omega_{n} - j\omega}{\zeta_{n}^{2}\omega_{n}^{2} + \omega^{2}}\;\; \Longrightarrow \;\;|V_{E}(\omega)|^{4} = \frac{|C_{n}|^{4}}{(\zeta_{n}^{2}\omega_{n}^{2} + \omega^{2})^{2}}.
\end{equation}
Then, we evaluate the numerator and denominator of Eq.~\eqref{sample:RMSBW} to obtain
\begin{equation}
\begin{aligned}
\displaystyle\int_{0}^{\infty}\omega^{2}|V_{E}(\omega)|^{4}d\omega &= |C_{n}|^{4}\Bigg(\frac{1}{\zeta_{n}\omega_{n}}\tan^{-1}\Big(\frac{\omega}{\zeta_{n}\omega_{n}}\Big)\Biggr|_{0}^{\infty} - \Bigg( \frac{1}{2(\zeta_{n}\omega_{n})}\tan^{-1}\Big(\frac{\omega}{\zeta_{n}\omega_{n}}\Big) + \frac{\omega}{2((\zeta_{n}\omega_{n})^{2}+\omega^{2})}    \Bigg)\Biggr|_{0}^{\infty}\Bigg)\\
&= |C_{n}|^{4}\Bigg( \Big(\frac{\pi}{2\zeta_{n}\omega_{n}}\Big) - \Big(\frac{\pi}{4\zeta_{n}\omega_{n}}\Big) - \lim_{\omega \to \infty}\Big(\frac{1}{2}\Big(\frac{\zeta_{n}\omega_{n})^{2}}{\omega} + \omega\Big)^{-1}\Big)\Bigg)\\
&= |C_{n}|^{4}\frac{\pi}{4}\Big(\frac{1}{\zeta_{n}\omega_{n}}\Big).
\end{aligned}
\end{equation}
We evaluate the denominator in a similar manner,
\begin{equation}
\begin{aligned}
\displaystyle\int_{0}^{\infty}|V_{E}(\omega)|^{4}d\omega &= |C_{n}|^{4}\Bigg( \frac{1}{2(\zeta_{n}\omega_{n})^{3}}\tan^{-1}\Big(\frac{\omega}{\zeta_{n}\omega_{n}}\Big) + \frac{\omega}{2(\zeta_{n}\omega_{n})^{2}((\zeta_{n}\omega_{n})^{2}+\omega^{2})}    \Bigg)\Biggr|_{0}^{\infty}\\
&= |C_{n}|^{4}\Bigg( \Big(\frac{\pi}{2\zeta_{n}^{3}\omega_{n}^{3}}\Big) - \lim_{\omega \to \infty}\Big(\frac{(\zeta_{n}\omega_{n})^{2}}{2}\Big(\frac{\zeta_{n}\omega_{n})^{2}}{\omega} + \omega\Big)^{-1}\Big)\Bigg)\\
&= |C_{n}|^{4}\frac{\pi}{4}\Big(\frac{1}{\zeta_{n}^{3}\omega_{n}^{3}}\Big).
\end{aligned}
\end{equation}
Substitute these quantities into Eq.~\eqref{sample:RMSBW} to obtain the ARMS Bandwidth, which recovers the classical bandwidth expression for a lightly damped, SDOF linear time-invariant oscillator (or single mode):
\begin{equation}
\Delta\omega_{\text{ARMS}} = 2\sqrt{\frac{\frac{\pi}{4}\frac{1}{\zeta_{n}\omega_{n}}}{\frac{\pi}{4}\frac{1}{\zeta_{n}^{3}\omega_{n}^{3}}}}= 2\zeta_{n}\omega_{n}.
\end{equation}

\subsection{Wavelet Transforms, Wavelet Transform Normalizations, and Methods for Modal Participation Extraction} 
The selection of wavelet transform processing parameters, namely normalization, can affect the quality of the ARMS Bandwidth computation. This section provides additional analysis and discussion concerning the appropriate wavelet transform normalization to ensure accuracy of the modal participations extracted in the data-driven methodology. In addition, different wavelet transform implementations and modal participation extraction methods are investigated to determine their effects on ARMS Bandwidth estimates.

Most wavelet transform implementations use either $L_{1}$- or $L_{2}$-normalization, with the former preserving amplitudes at different scales and the latter preserving energy. For our standard data-driven methodology, $L_{1}$-normalization is necessary to extract accurate modal amplitudes (modal participations) from a maximum wavelet transform plot to obtain an accurate ARMS Bandwidth estimate. To demonstrate this, a study of ARMS Bandwidth variation of the two-DOF system described in Section \ref{twoDOFsection} with new system parameters under different normalizations, wavelet transform implementations, and modal participation extraction methods was performed. Mass one was set to 1, mass two set to 1.2, the stiffnesses $k_{1}$, $k_{2}$, and $k_{3}$ set to 1000, 2000, and 3500, respectively, and the damping coefficients $c_{1}$, $c_{2}$, and $c_{3}$ set to 2, 1, and 1, respectively. Solving the polynomial eigenvalue problem, $(\matr{M}\lambda^{2} + \matr{C}\lambda + \matr{K})\vect{\phi}=0$, we obtain an in-phase mode and out-of-phase mode with modal properties $(\omega_{1},\zeta_{1},\vect{\phi}_{1})=(42.5 \;\text{rad\slash s},0.0207,[0.858 \;0.514]^{T})$ and $(\omega_{2},\zeta_{2},\vect{\phi}_{2})=(76 \;\text{rad\slash s},0.0191,[0.584;-0.812]^{T})$, respectively. Each mass was excited using a pulse of the same form as Eq.~\eqref{sample:Impulse} and each resulting velocity signal $v_{ij}$, where $i$ is the measurement DOF and $j$ is the excitation DOF, was used to compute ARMS Bandwidth.

The standard data-driven methodology discussed in Section \ref{datadrivenmethod}, i.e., using the Morlet wavelet transform to generate the maximum wavelet transform, was used with $L_{1}$- and $L_{2}$-normalized wavelets to generate ARMS Bandwidth estimates, which were compared to exact ARMS Bandwidth values using numerically-exact modal participations derived from simulated modal responses and numerical eigenvectors. Furthermore, MATLAB's native continuous analytic Morse wavelet transform ($L_{1}$-normalization) function \emph{cwt($\cdot$)} and synchro-squeezed analytic Morlet wavelet transform ($L_{1}$-normalization) function \emph{wsst($\cdot$)} were assessed. An additional method for extracting modal participations using the inverse wavelet transform (\emph{icwt($\cdot$)} and \emph{iwsst($\cdot$)}) was evaluated. In this method, the wavelet transform of a signal is processed, harmonic regions associated with modal responses are isolated and converted back into isolated time histories via the inverse wavelet transform, and the maximum of each harmonic component time history is selected as the modal participation for each mode. A final method for extracting modal participations using FRFs was also evaluated. In this method, modal curve-fits are generated for a set of FRFs to obtain the residues, i.e., modal participations, directly via $R_{ijn}=Q_{n}\phi_{in}\phi_{jn}$ where $R_{ijn}$ is the residue of mode $n$ at measurement DOF $i$ due to excitation at DOF $j$, $Q_{n}$ is the complex modal scaling coefficient, and $\phi_{in}$ and $\phi_{jn}$ are the modeshape entries associated with mode $n$ and degrees-of-freedom $i$ and $j$. A summary of ARMS Bandwidth is provided in Table \ref{tab:summary}.

\begin{table}[H]
\caption{\label{tab:summary} Summary of analysis results; error was computed using numerically-exact results.}
\centering
\begin{tabular}{lccccc}
\hline
\multicolumn{1}{l}{\multirow{2}{*}{Analysis Types}} &
 \multicolumn{4}{c}{ARMS Bandwidth} & 
 \multicolumn{1}{l}{\multirow{2}{*}{Mean Error}} \\ \cline{2-5}
\multicolumn{1}{c}{} & 
\multicolumn{1}{c}{$v_{11}$} & 
\multicolumn{1}{c}{$v_{12}$} & 
\multicolumn{1}{c}{$v_{21}$} & 
\multicolumn{1}{c}{$v_{22}$} & 
\multicolumn{1}{c}{}\\ \hline
\multicolumn{6}{c}{\emph{Standard Data-driven Methodology}} \\
Numerically-Exact & 1.992 & 2.171 & 2.177 & 2.435 & --\\
Standard Morlet Wavelet Transform ($L_{1}$) & 1.998 & 2.204 & 2.204 & 2.453 & 0.96\%\\
Standard Morlet Wavelet Transform ($L_{2}$) & 1.946 & 2.127 & 2.127 & 2.368 & -2.35\%\\
MATLAB's Continuous Wavelet Transform ($L_{1}$) & 1.992 & 2.191 & 2.191 & 2.430 & 0.35\%\\
MATLAB's Synchro-squeezed Wavelet Transform ($L_{1}$) & 1.978 & 2.222 & 2.222 & 2.497 & 1.58\%\\
\multicolumn{6}{c}{\emph{Inverse Wavelet Transform Methodology}}\\
MATLAB's Continuous Wavelet Transform ($L_{1}$) & 2.040 & 2.192 & 2.192 & 2.428 & 0.95\%\\
MATLAB's Synchro-squeezed Wavelet Transform ($L_{1}$) & 2.000 & 2.208 & 2.208 & 2.442 & 0.96\%\\
\multicolumn{6}{c}{\emph{Frequency Response Function Methodology}}\\
Least-squares Complex Exponential Curve-fit & 1.993 & 2.171 & 2.191 & 2.435 & 0.17\%\\
\hline
\end{tabular}
\end{table}

All methods and analysis types yield ARMS Bandwidth estimates within 2.5\% of the numerically-exact values. However, the error drops off significantly when $L_{1}$-normalization is used with the continuous wavelet transform. Both the standard data-driven and inverse wavelet transform methodologies provide similar ARMS Bandwidth estimates while the FRF methodology provides the best estimate. However, the FRF methodology requires clean, linear responses, a high-quality curve-fit, and the ability to directly obtain the residues. Achieving these requirements is oftentimes challenging with real-world modal test data and industrial black-box modal analysis software suites.

It is clear that all modal participation extraction methodologies are sufficient to obtain accurate ARMS Bandwidth estimates and that $L_{1}$-normalization results in more accurate estimates compared to $L_{2}$-normalization results. Furthermore, this study demonstrates that accurate modal participations can be obtained directly using residues from curve-fits of existing or new FRF data from modal tests.

\subsection{The Use of Roving Hammer and Roving Sensor Data for the Computation of Spatial ARMS Bandwidth} 
When using modal test data to compute the spatial distribution of local ARMS Bandwidth, henceforth referred to as spatial bandwidth, one may opt to use roving hammer or roving sensor data depending on sensor and data acquisition system constraints, mass loading concerns, test expediency, and/or available data. The utility and interpretation of spatial bandwidth results depends greatly on the data set used in its generation and, thus, careful consideration  is required prior to acquiring or selecting modal test data sets.

\begin{figure}[H]
\centering
\includegraphics[width=1\textwidth]{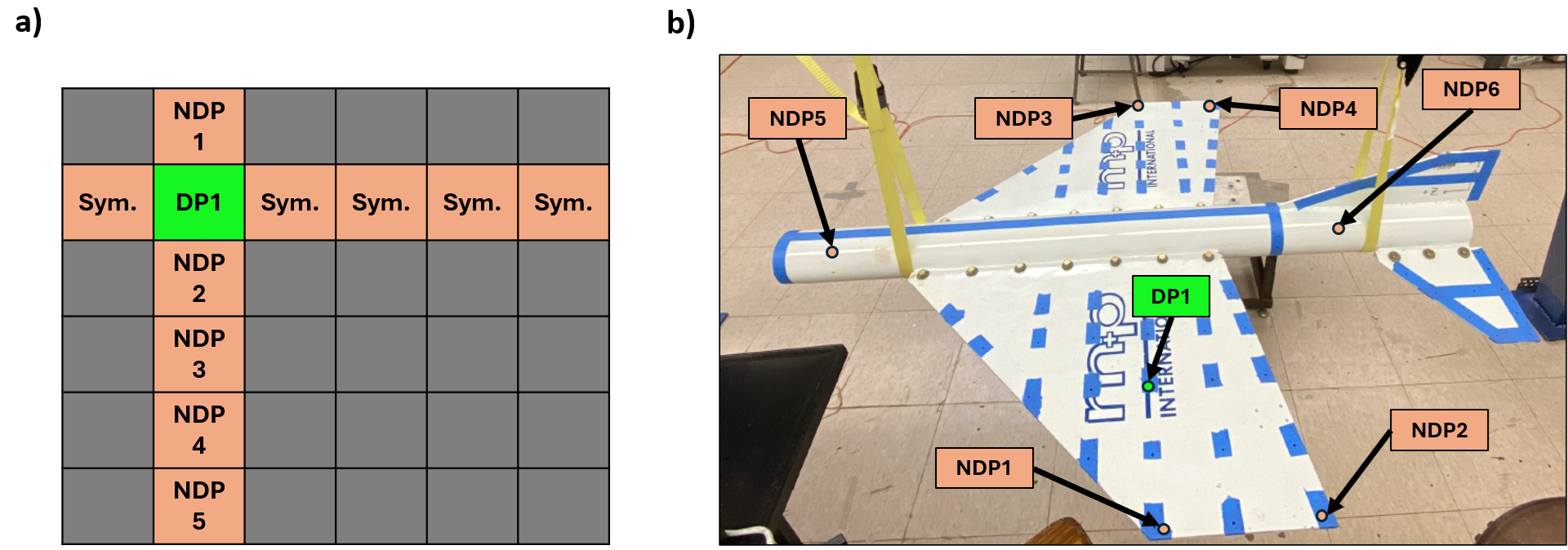}
\caption{Roving hammer modal test scenario: a) input-output matrix and b) physical model with the drive point and non-drive points identified.}
\label{fig:rovinghammer}
\end{figure}

Presented in Fig. \ref{fig:rovinghammer} is a visual representation of the input-output matrix associated with a roving hammer modal test of the linear, proportionally-damped aircraft model. Sensor locations are indicated by non-drive points (NDPs) and the excitation point (also measured) is indicated by the drive point (DP) on the right of Fig. \ref{fig:rovinghammer}. This is an idealized test setup to characterize the spatial dissipative capacity of the airframe under an impulsive load, e.g., when an aircraft ejects a wing store. An impulse is applied to DP1 and the input-output matrix is populated along a column; the input-output matrix is symmetric due to reciprocity and, thus, the transposed column entries are also populated. Computing the spatial bandwidth distribution for DP1 and all NDPs results in a spatial bandwidth plot similar in form to Fig. \ref{fig:bwmap} that ultimately describes how each location dissipates energy due to impulsive excitation at DP1. This information can be used for determining the optimal location for vibration mitigation and avoidance. In contrast, a roving sensor modal test, assuming a single sensor location, would only populate input-output matrix diagonal entries and would produce a disjointed spatial bandwidth plot. Unlike in the roving hammer case, this spatial bandwidth plot stitches together the local dissipative capacity at a point due to excitation at that same point for all drive points. From a practical standpoint, drive points for which local ARMS Bandwidth are minima could be candidates for ground vibration test excitation points. On the contrary, drive points for which local ARMS Bandwidth are maxima could inform candidate locations for sensitive equipment. While this information is useful, it does not provide the same utility for the purpose of the test as the spatial bandwidth plot resulting from a roving hammer modal test.

When determining the set of modal test data to acquire of use to generate spatial bandwidth plots, one must consider the intended use of the results.
Spatial bandwidth for vibration mitigation and avoidance is most useful when computed at the DP and all NDPs across a structure under impulsive excitation at a single point along a single direction, e.g., when using roving hammer modal test data, as it quantifies the structure's spatial dissipative capacity due to a local energy source, such as the ejection of a wing store. Spatial bandwidth using only DP data generates a disjointed view of the structure's dissipative capacity and provides fewer use-cases.

\section*{Funding Sources}
Benjamin J. Chang was supported in part by the Naval Air Warfare Center Aircraft Division (NAWCAD), Naval Innovative Science and Engineering (NISE) Program. Any opinions, findings, and conclusions or recommendations expressed in this work are those of the authors and do not necessarily reflect the views of NAWCAD.

Keegan J. Moore was supported in part by the Air Force Office of Scientific Research Young Investigator Program under grant number FA9550-22-1-0295.

%\section*{Acknowledgments}

\bibliography{sample}

\end{document}